\documentclass[fleqn,usenatbib]{mnras}

\usepackage{newtxtext,newtxmath}

\usepackage[T1]{fontenc}
\usepackage{orcidlink}

\DeclareRobustCommand{\VAN}[3]{#2}
\let\VANthebibliography\thebibliography
\def\thebibliography{\DeclareRobustCommand{\VAN}[3]{##3}\VANthebibliography}

\usepackage{graphicx}	
\usepackage{amsmath}	

\title[z=6.63 NIRCam-dark DSFG]{An almost NIRCam-dark dusty star-forming galaxy at z=6.63}

\author[L. Bing et al.]{Longji Bing,$^{1}$\thanks{E-mail: l.bing@sussex.ac.uk}\orcidlink{0000-0002-0440-7129}
Seb Oliver,$^{1}$
Mengyuan Xiao,$^{2}$
Guilaine Lagache,$^{3}$
Sylvia Adscheid,$^{4}$
Daizhong Liu,$^{5}$\orcidlink{0000-0001-9773-7479}
\newauthor{Benjamin Magnelli,$^{6}$}
Roberto Neri,$^{7}$
Miroslava Dessauges-Zavadsky,$^{2}$
Anton M. Koekemoer,$^{8}$
\newauthor{Maximilien Franco,$^{6}$}
Shuowen Jin,$^{9,10}$
Olivia R. Cooper,$^{11,12}$\orcidlink{0000-0003-3881-1397}
Andreas L. Faisst,$^{13}$\orcidlink{0000-0002-9382-9832}
Catilin M. Casey,$^{14}$
\newauthor{Jeyhan S. Kartaltepe,$^{15}$}
Hollis Akins,$^{16}$
Alexandre Beelen,$^{3}$
David Elbaz,$^{6}$
Steven Gillman,$^{9,10}$
\newauthor{Santosh Harish,$^{8}$}
Arianna S. Long,$^{17}$\orcidlink{0000-0002-7530-8857}
Henry Joy McCracken,$^{18}$
Pascal Oesch,$^{2}$
Louise Paquereau,$^{18}$
\newauthor{Nicolas Ponthieu,$^{19}$}
Jason Rhodes,$^{20}$
Brant Robertson,$^{21}$
David B. Sanders,$^{22}$\orcidlink{0000-0002-1233-9998}
Marko Shuntov,$^{9,23}$
\newauthor{Stephen Wilkins$^{1}$}
\\
Affiliations can be found after the references.
}

\date{Accepted XXX. Received YYY; in original form ZZZ}

\pubyear{\the\year{}}

\begin{document}
\label{firstpage}
\pagerange{\pageref{firstpage}--\pageref{lastpage}}
\maketitle

\begin{abstract}
We present AC-2168, an almost NIRCam-dark, millimetre-bright galaxy in the COSMOS field. The source was identified blindly in ALMA Band-4 continuum data and remains undetected in the COSMOS-Web DR1 NIRCam catalogue. We spectroscopically confirm a redshift of $z_{\rm spec}=6.631$ from [\ion{C}{ii}] 158\,$\mu$m and four tentatively detected CO lines in NOEMA and ALMA data. SED fitting to near-IR to millimetre photometry yields $\rm L_{IR}=1.6\times10^{12}\,L_\odot$, an SFR of $\rm 244\,M_\odot/yr$, heavy dust attenuation $\rm A_V=5.4$ mag, and a stellar mass $\rm M_\star=3.7\times10^{10}\,M_\odot$. From the millimetre continuum and [\ion{C}{ii}] emission, we infer a warm ISM with $\rm T_{\rm dust}=60K$, $\rm M_{dust}=3.0\times10^{8}\,M_\odot$ and $\rm M_{gas}=4.1\times10^{10}\,M_\odot$. AC-2168 has a gas fraction ($f_{\rm gas}=M_{\rm gas}/(M_\star+M_{\rm gas})$) of $\simeq52\%$, a short depletion time of $\rm \sim170Myr$, a compact ($\rm \sim1kpc$) dust-continuum size, and an SFR consistent with the star-forming main sequence at its mass. These properties match expectations for progenitors of massive quiescent galaxies at the peak of their assembly, as implied by NIRSpec-based SFHs of $z\sim4-5$ systems. Using the blind detection, we estimate a space density of $\rm 7.8^{+18.0}_{-6.5}\times10^{-6}\,cMpc^{-3}$ for AC-2168-like NIRCam-dark galaxies at $z\sim6-7$, $\sim42\%$ of the abundance of massive quiescent galaxies at $z\sim4-5$. No overdensity of Ly$\alpha$ emitters or Lyman-break galaxies is found nearby, suggesting AC-2168 does not lie in a prominent protocluster and highlighting the importance of unbiased blind surveys for this population.

\end{abstract}




\section{Introduction}

The emergence of massive galaxies within the first few gigayears (Gyrs) after the Big Bang, either active in star formation \citep[e.g.,][]{Ferrara+23,Xiao+24} or quiesence \citep[e.g.,][]{Carnall+23,Nanayakkara+24,deGraaff+25,Xiao+25, Weibel+25}, imply that they must undergo, or have experienced,  brief but highly efficient phases of mass assembly. Theoretically, dissipative "wet" compaction of gas-rich galaxies and the associated central gas pile-up can trigger a short, intense burst that exhausts/expels fuel and inaugurates inside-out quenching and maintains the quiescence when halos are sufficiently massive \citep[e.g.,][]{Dekel+14,Zolotov+15,Tacchella+15,Lapiner+23}. {\em JWST} spectroscopy now confirms quiescent galaxies (QGs) at $z\sim4-5$ with stellar populations that formed quickly at $z\gtrsim6-8$ on $\sim$100-300\,Myr timescales \citep[e.g.,][]{Valentino+23,Carnall+24,Nanayakkara+24,deGraaff+25}, sharpening the need to identify their progenitors. 

(Sub)millimetre blind surveys have established a population of dusty star-forming galaxies (DSFGs) with ${\rm SFR} \sim 10^{2}$-$10^{3}\,{\rm M_\odot\,yr^{-1}}$ that are heavily attenuated in the rest-UV/optical and dominate the high luminosity tail of the infrared luminosity function (IRLF) at high redshift \citep{Casey+14,Hodge+20}. The most obscured population of DSFGs invisible in the deepest \emph{HST} data but bright in mid- and far-IR, also known as the "optically/NIR dark" or H-dropout galaxies, could contribute to $\sim$20\% of ALMA blind detections \citep{Franco+18} and dominate the population of massive star-forming galaxies at $z\gtrsim4$ \citep{Wang+19, Xiao+23,Manning+25}. These include the prototype SMG \mbox{HDF\,850.1} at $z=5.183$ \citep{Walter+12}, which remained undetected in the rest-frame UV/optical before the sensitive JWST NIRCam observation was carried out \citep{Sun+24,Herard-Demanche+25,Lagache+25}. In addition, these observations also revealed a large-scale overdensity of star-forming galaxies associated with \mbox{HDF\,850.1}.


At $z \gtrsim 6$, most securely detected DSFGs are drawn from samples with complex selection functions, complicating the interpretation of their intrinsic abundances and environments. These include the hyper-luminous DSFGs HFLS3 at $z = 6.34$ \citep{Riechers+13} and the SPT\,0311-58 system at $z=6.90$ \citep{Marrone+18} identified from the {\em Herschel} HERMES \citep{Oliver+12} and SPT-SZ survey \citep{Mocanu+13}. Although still intrinsically luminous, high-resolution follow-up observations revealed that most of these sources are lensed by massive galaxies in the foreground. Apart from lensed sources, other z$>$6 DSFGs are mostly identified from targeted observations on high-z sources. For example, ALMA targeted observations on luminous quasars frequently detect companion galaxies in the primary beams with [\ion{C}{ii}] at a similar redshift to the quasars \citep{Decarli+17,Venemans+20,Pensabene+21}. Targeted pointings toward UV-selected galaxies at $z \sim 7$ have also revealed dust-obscured [\ion{C}{ii}]-emitting companions and imply a substantial (10-25\%) contribution of obscured star formation to the total SFR density at $z\sim7$ \citep{Fudamoto+21}. The proximity of these DSFGs in position and redshift to massive galaxies and quasars, however, indicates that their selection is biased to the most massive haloes.

Unbiased blind millimetre selection in blank fields provides a complementary route to probe the obscured phase of galaxy assembly and enables volume-anchored statistics. In particular, blind ALMA \citep[e.g.,][]{Franco+18,Gomez-Guijarro+22a,Casey+21,Zavala+21,Long+24b} and single-dish surveys \citep[e.g.,][]{Weiss+09,Geach+17,Simpson+19,Bing+23,Gao+24} with both wide area coverage and adequately deep observation surface high-$z$ DSFGs and begin to place number-density constraints, revealing a substantial number of $z > 4$ sources or candidates \citep[e.g.,][]{Danielson+17,Jin+19,Dudzeviciute+21,Mitsuhashi+21,Jin+22,Chen+22,Berta+25,Bethermin+25}. Nevertheless, spectroscopic confirmation beyond $z > 6$ among these blindly detected millimetre sources remains scarce, where only one of them (GNz7q) is securely confirmed to be at $z_{\rm spec}=7.19$ (\citealt{Fujimoto+22}, see also \citealt{Berta+25}). These results emphasise that even a single blind, unlensed detection at $z > 6$ is particularly valuable: it avoids lensing priors and intended pointings on massive galaxies and quasars, clarifies typical environments, and allows a direct estimate of the space density once survey volume and completeness are accounted for.

{\emph{JWST}} now exposes more details of these infrared-bright, optically/NIR faint systems and reveals an even more extreme population of sources that remain faint or undetected even in deep NIRCam imaging ("NIRCam-dark"). Early JWST results have clarified the demographics of optically/NIR-dark systems at $z \sim 2-8$ that were missed by rest-frame UV selections, while also underscoring that the most obscured subsamples can drop beneath NIRCam depths at $\rm \gtrsim2\mu m$ \citep[e.g.,][]{Barrufet+23a,Mckinney+23,Perez-Gonzalez+24,Fujimoto+25,Manning+25}. In particular, observations by the ASPIRE survey \citep{Wang+23,Yang+23} have uncovered two NIRCam-dark starbursts spectroscopically confirmed to be within the reionisation era. Residing at $z_{\rm spec} \simeq 6.6$, these two sources are identified as quasar companions with robust ALMA continuum and [\ion{C}{ii}] detections but NIRCam non-detections ($m_{\rm AB} > 28$ at 3-5\,$\mu$m) \citep{Sun+25}. They argue that such systems may possibly link to the $z \gtrsim 4$ QGs through the qualitative match between the inferred SFR and $M_*$ of their sample and the star formation history (SFH) of QG progenitors at the same redshift as their sample. 

The selection in the proximity of high-z quasars leads to large and model-dependent uncertainty on the abundance of these JWST/NIRCam-dark galaxies, as well as the connection between environment and their obscured rapid assembly and metal enrichment at such an early time. This calls for unbiased searching on this population for proper demographics and understanding the physical driver of their formation. In light of these motivations, we present AC-2168, a NIRCam-dark DSFG at $z=6.631$ in the COSMOS field initially identified by ALMA and spectroscopically confirmed with NOEMA. Despite its ultra-luminous infrared luminosity of $\rm >10^{12}L_\odot$, its rest-frame UV emission is completely hidden and lacks a counterpart in even the deepest source catalogue in the COSMOS field \citep{Casey+23,Shuntov+25} based on NIRCam images, indicating exceptional dust obscuration. Moreover, unlike most known $z > 6$ dusty systems, our discovery arises from a serendipitous detection uncorrelated with the main target in redshift.  This leads to a model-independent, volume-based number density estimation of NIRCam-dark galaxies and abundance comparison with other galaxy populations, as well as enabling an environmentally unbiased view on one of the highest redshift DSFGs. 

We summarise the data and identification of AC-2168 in Section~\ref{sect:data_and_red}, and present the spectroscopic confirmation and physical properties of it in Section~\ref{sect:results}. In Section~\ref{sect:discussion},  we will discuss implications of AC-2168 for massive-galaxy evolution, focusing on star-formation histories and number-density consistencies with high-$z$ QGs, as well as the distinct lack of coincidence with galaxy overdensities. Throughout the paper, we adopt the best-fit cosmological parameters from the \citet{[Planck+18} with $\rm \Omega_M=0.310$ and $\rm H_0=67.7 km/s/Mpc$, and an initial mass function following \citet{Chabrier+03}.

%
\section{Source Identification, Observation and Data reduction}\label{sect:data_and_red}

We select our sample by identifying orphan millimetre sources in the v20220606 compilation of A3COSMOS \citep{Liu+19a,Adscheid+24}, which is covered by COSMOS-Web NIRCam imaging \citep{Casey+23} but remains undetected in its NIRCam source catalogue \citep{Shuntov+25}. These two datasets comprise the largest overlap in deep ALMA and JWST NIRCam observations in a single deep field. Given the expected scarcity of massive DSFGs in the early Universe, the wide coverage is optimal for our search. A3COSMOS contains 2207 blind detections from ALMA observations publicly released by 2022-06-06, covering 896 arcmin$^2$ in total with Bands 3-10. 59\% of them fall into the footprint of COSMOS-Web, the largest JWST blank deep-field survey. The NIRCam observation of COSMOS-Web covers 0.54 deg$^2$ nearly continuously in the COSMOS field, with point source sensitivity of 26.9-28.3 mag at 5$\sigma$ in the F115W, F150W, F277W and F444W bands \citep{Casey+23,Franco+25,Harish+25,Shuntov+25}. In this section, we will describe the method used to identify AC-2168 from A3COSMOS and COSMOS-Web data and catalogues and summarise the exact archival ALMA and JWST datasets and NOEMA follow-up observations that are applied in the redshift confirmation and source property measurements. 

\subsection{Identification of the source}\label{sbsc:sourceid}

\begin{figure*}
    \centering
    \includegraphics[width=1.0\linewidth]{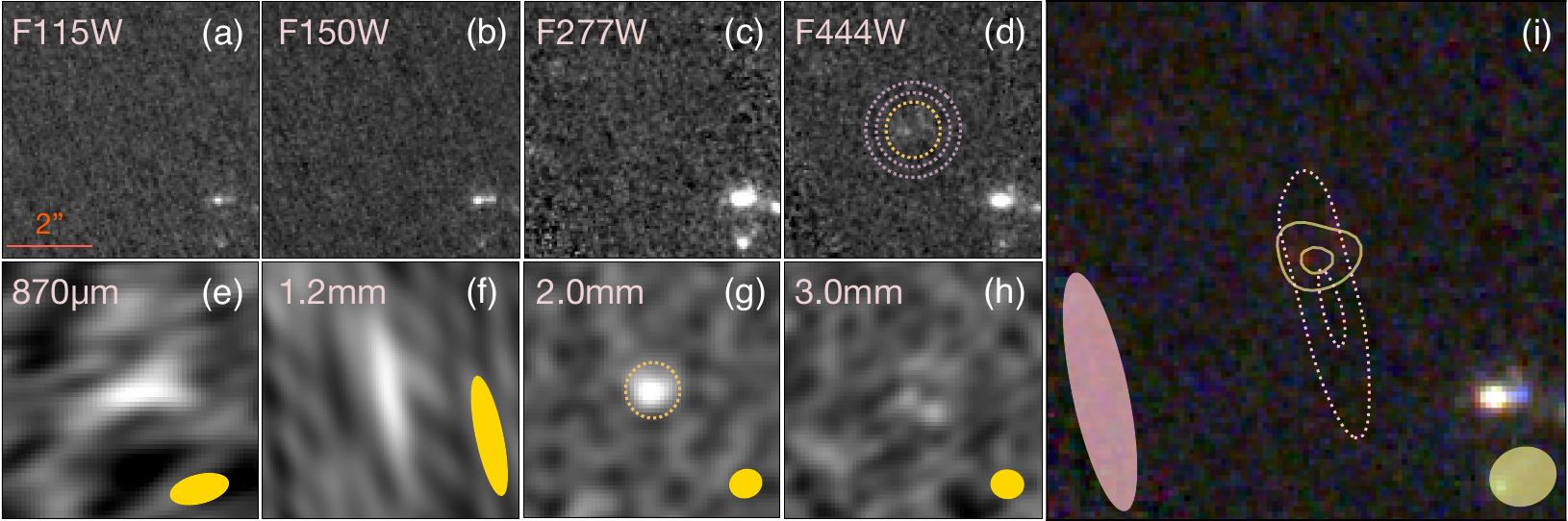}
    \caption{Near-IR to millimetre images of 6"$\times$6" size centred on AC-2168. (a)-(d): JWST NIRCam images from the COSMOS-Web survey. The aperture and annulus used in NIRCam photometry are outlined by orange and purple dotted circles, respectively (e)-(h): (sub) millimetre continuum data from ALMA archival data (870$\mu$m, 2.0mm and 3.0mm) and NOEMA DDT observation (1.2mm). The beam sizes are shown as yellow ellipses in the bottom right of the corresponding panels. The aperture used to extract ALMA Band 3+4 spectra is outlined by the yellow dotted circle. (i): ALMA 2mm continuum and NOEMA [\ion{C}{II}] S/N contour overlaid on the pseudo-colour image generated by NIRCam F150W (blue), F277W (green) and F444W (red) data. The contour levels of the 2\,mm continuum (yellow) and [\ion{C}{II}] (pink) correspond to S/N=5 and 10 and S/N=3 and 5, respectively. Their beam shapes are shown as a yellow and pink ellipse in the bottom right.}
    \label{fig:cutout}
\end{figure*}


We summarise the stepwise identification of our sample using A3COSMOS and COSMOS-Web data as follows. First, we crossmatch all blindly detected A3COSMOS sources in the COSMOS-Web NIRCam coverage with the \texttt{SExtractor++} source catalogue in COSMOS-Web DR1 \citep{Shuntov+25} to eliminate those with a counterpart. The matching radii are set to 1" to account for the resolution of ALMA continuum data, the size of most extended high-z galaxies \citep[e.g.][]{} and the possible offset between dust and stellar emission in galaxies \citep{}. All matched sources are significantly detected at $\rm S/N\ge5$ in at least the two NIRCam bands of the longest wavelength available in COSMOS-Web (F277W and F444W). We then perform a visual inspection of the NIRCam images at the location of each unmatched A3COSMOS source to identify those whose non-detection in NIRCam arises from contamination by bright nearby stars or galaxies. The remaining sources are further inspected in the ALMA archive to check for observations that are comparably deep or more sensitive than the images used in the blind detection in A3COSMOS. The majority of these NIRCam-dark A3COSMOS sources show no counterparts in deeper or comparably deep ALMA data at $\rm S/N\gtrsim3$, suggesting that they are spurious ALMA sources. 

Having eliminated bright source contamination in NIRCam and likely spurious detections in A3COSMOS, a sample of four blindly detected ALMA continuum sources remains undetected in the COSMOS-Web DR1 catalogue. Among these sources, AC-2168 (RA: 10:00:48.9, Dec: +02:30:03.8) stands out as the only blind serendipitous detection considered as highly confident \citep{Adscheid+24}, with S/N$\gtrsim$5.4 at the peak in the A3COSMOS continuum image. Such a high S/N detection is unlikely to be a random noise spike, with a spurious rate $\lesssim8\%$ \citep{Liu+19a}. Moreover, the source is detected at the same position and with a similar S/N in the A3COSMOS-reduced continuum from four consecutive ALMA Band 4 tunings. The high S/N and consistency of detection in different datasets strongly support its astrophysical origin, making it the most secure NIRCam-dark source within the current A3COSMOS and COSMOS-Web coverage. AC-2168 is also the only 2\,mm (Band 4) blindly detected source with no NIRCam counterparts in COSMOS-Web coverage. Fig.~\ref{fig:cutout} shows the multiwavelength images cutout centred on AC-2168, illustrating its faintness in the NIRCam image.

\subsection{Archival and New Millimeter Observations}\label{sbsc:almaobs}

We extract the spectral cube of ALMA observations of AC-2168 from the reduced data in the ALMA Science Archive from project 2019.1.01722.S (PI: W.-H. Wang). With four tunings at each band, the Band 3 and Band 4 spectral cubes from the observations continuously cover the spectrum within 85.086-108.258\, GHz and 131.189-154.264\, GHz. These ancillary observations centred on another DSFG at RA: 10:00:48.71 Dec: 02:30:17.91, which has \texttt{LePHARE} $z_{\rm phot}\sim3$ based on the COSMOS-Web DR1 \texttt{SExtractor++} photometry \citep{Arnouts+11, Shuntov+25}. The observations in Band 3 and Band 4 were carried out with C43-5 and C43-4 configurations, resulting in beam sizes of 0.79\arcsec$\times$0.69\arcsec and 0.79\arcsec$\times$0.68\arcsec, respectively. For the analysis on the continuum data, in addition to the image from A3COSMOS, we further stack the visibilities of the continuum within each band using the calibrated measurement sets with CASA 6.7.0, which aims to maximise the continuum sensitivity and S/N to benefit the flux and size measurement. We measured the 2.0\ mm and 3.0\ mm fluxes, or upper limits, by fitting these visibilities using {\sc uvmodelfit} in {\sc CASA} and a Gaussian source model with prior positions at the continuum position of the A3COSMOS detection. The continuum size is also constrained simultaneously under the assumption of a circular Gaussian model, which determines an FWHM of 0.21$\pm$0.09" at 2.0\,mm. In addition to the Band 3 and Band 4 observations, AC-2168 is also covered by the Band 7 observation of project 2013.1.00884.S (PI: D. Alexander) at the edge of the primary beam. The source flux at 870$\mu$m is measured using the same method as we applied to the 2.0\,mm and 3.0\,mm continuum. 

Using the archival ALMA Band 3 and Band 4 spectra extracted at the source position, we identify a possible redshift solution of AC-2168 using the method described in \citet{Bing+24}, which relies on the joint information of source photometric redshift, infrared luminosity, and millimetre spectrum (see Sect.~\ref{sbsc:zspec} for the identified CO lines). The analysis picks up four tentatively detected CO lines in the Band 3 and Band 4 ALMA spectra (see Sect.~\ref{sbsc:zspec}), corresponding to a redshift at $z=6.635\pm0.005$. To robustly confirm or exclude this redshift solution, a follow-up observation on AC-2168 was performed with NOEMA Band 3 (1\,mm) at the C configuration under project E24AI (PI: Bing) in April 2025. The LO frequency of the PolyFiX correlator is set to 243\,GHz to cover the spectrum within 231.384-239.128\,GHz and 246.872-254.616\,GHz. The total on-source observing time is 2.3 hours. The NOEMA observation is first calibrated using CLIC and imaged by MAPPING under GILDAS. We use 3C273 for bandpass calibration, J1028-0236 for phase calibration and 2010+723 for flux calibration. With the calibrated data, we generate the uv table with the original resolution of 2 MHz. We also produce the continuum uv table of each source by directly compressing all corresponding lower sideband (LSB) and upper sideband (USB) data with MAPPING and producing a cleaned image at 1.2\,mm (see Fig.~\ref{fig:cutout}), with a resulting beam size of 2.87"$\times$0.64". The continuum fluxes are measured by fitting the visibility using the uvfit function in MAPPING with a point source model, taking into account the larger and elongated beam compared to ALMA Band 4.

AC-2168 also falls into the footprint of the STUDIES survey, which observed the COSMOS region within the CANDELS survey \citep{Grogin+11,Koekemoer+11} with JCMT SCUBA2 at 850$\mu$m and 450$\mu$m. We refer the readers to \citet{Wang+17} and \citet{Gao+24} for the details of the survey design, data reduction, and source catalogue of STUDIES. We crossmatch AC-2168 with the STUDIES source catalogue with a diameter of 6.5", corresponding to the FWHM of the SCUBA2 850$\mu$m beam. A S/N$\sim$4 850$\mu$m source is identified. However, no counterpart is found in the 450$\mu$m map. This non-detection at the ALMA coordinate in the STUDIES 450$\mu$m map results in a 2$\sigma$ flux upper limit at 2.6\,mJy when considering both instrumental and confusion noise at the position of AC-2168 \citep{Gao+24}. 

\subsection{Ancillary data}\label{sbsc:archiveobs}

AC-2168 is covered by JWST/NIRCam \citep{Rigby+23,Rieke+23} imaging data of the COSMOS-Web survey \citep{Casey+23} with F115W, F150W, F277W, and F444W filters. The details of NIRCam data reduction can be found in \citet{Franco+25}. The point source sensitivity of the NIRCam data around AC-2168 reaches $\sim$28 AB mag in F444W. Although COSMOS-Web also conducted MIRI observations over 0.2\,deg$^2$ in the COSMOS field in parallel to NIRCam, these mid-IR observations do not cover AC-2168.

As AC-2168 is not detected by the COSMOS-Web DR1 \texttt{SExtractor}++ catalogue, we perform custom aperture photometry at its position to more accurately constrain the near-IR fluxes, SED, and source properties. The custom photometry is conducted in the four NIRCam bands. As shown in Fig.~\ref{fig:cutout}, although COSMOS-Web DR1 reports no detection,  there is a faint and diffuse source visible in only F444W data coincident with the position of millimetre continuum detections. We thus use a circular aperture of 0.6" radius for near-IR flux measurements centred on the position of the ALMA continuum, with an external annulus of inner radius 0.8" and width 0.2" for background measurements. The photometric aperture is chosen to match the size at which the F444W flux starts to converge with the increase of its radius. The background annulus has the same area as the region enclosed by the photometric aperture. The location and sizes of the photometric aperture and background annulus are illustrated in Fig.~\ref{fig:cutout}d. The NIRCam flux upper limits are presented in Table~\ref{tab:photometry}.

AC-2168 also falls into the primary beam of VLA S-band observation from COSMOS-XS, the deepest radio observation in the COSMOS field at arc-second resolution \citep{Algera+20,VanderVlugt+21}. 
No radio source in the COSMOS-XS 3 GHz source catalogue \citep{VanderVlugt+21} is detected within 2" of the position of AC-2168. With the sensitivity of COSMOS-XS data at the centre of the primary beam and the 400" offset of AC-2168 from the centre of its pointing, we constrain the 2$\sigma$ upper limits of 3\,GHz flux of AC-2168 at 2$\mu$Jy under the point source assumption. 

The complete summary of multiwavelength fluxes of AC-2168 from near-IR to radio could be found in Table~\ref{tab:photometry}.

\begin{table}
	\centering
	\caption{Multiwavelength photometry on AC-2168}
	\label{tab:photometry}
	\begin{tabular}{lll} 
		\hline
		Band & Unit &Flux Density \\
		\hline
		NIRCam F115W & nJy & $<$105 \\
		NIRCam F150W & nJy & $<$94 \\
		NIRCam F277W & nJy & $<$22 \\
        NIRCam F444W & nJy & 119$\pm$19 \\
        SCUBA2 450$\mu$m & mJy & $<$2.6 \\
        SCUBA2 850$\mu$m & mJy & 1.5$\pm$0.6 \\
        ALMA 870$\mu$m & mJy & 1.8$\pm$0.4 \\
        NOEMA 1.2mm & mJy & 0.52$\pm$0.14 \\
        ALMA 2.0mm & mJy & 0.20$\pm$0.02 \\
        ALMA 3.0mm & mJy & 0.037$\pm$0.015 \\
        VLA 3\,GHz & $\mu$Jy & $<$2.0 \\
		\hline
	\end{tabular}
\end{table}

\section{Results}\label{sect:results}

\subsection{Redshift of AC-2168}\label{sbsc:zspec}

\begin{figure}
    \centering
    \includegraphics[width=1\linewidth]{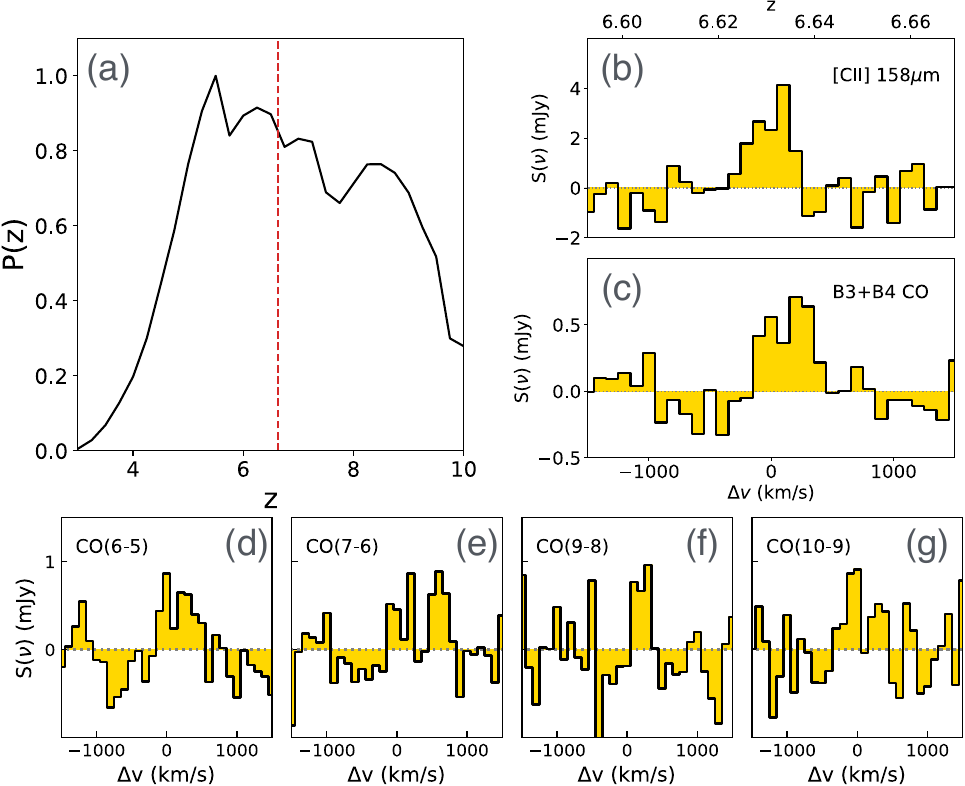}
    \caption{Data confirming the spectroscopic redshift of AC-2168. (a): The probability distribution of photometric redshift of AC-2168 from \texttt{CIGALE} fitting compared to the $z_{\rm spec}$ from ALMA and NOEMA spectra. (b): NOEMA 1.2mm spectrum of AC-2168 centered on [\ion{C}{II}]. (c): Stacked CO spectrum of AC-2168 based on ALMA Band 3 and Band 4 data. (d)-(g): ALMA Band 3 and Band 4 spectra centred on four individual CO lines within their coverage at the $z_{\rm spec}$ of AC-2168.}
    \label{fig:redshift}
\end{figure}

We estimate the photometric redshift of AC-2168 using the multiwavelength fluxes or upper limits in Table~\ref{tab:photometry}. The SED of AC-2168 is modelled by \texttt{CIGALE} \citep{Boquien+19,Yang+22}, a parametric SED fitting software based on an energy balance assumption. For the stellar population, we use the simple stellar population model from \citep{BC03} at solar metallicity, considering the presence of dust, and adopt a delayed-$\tau$ SFH with $\rm SFR(t) \propto$ $te^{(-t/\tau)}$ plus an optional recent starburst of 30 Myr in age. The age (\texttt{sfh.age\_main}), $\tau$ (\texttt{sfh.tau\_main}) and the fraction of the stellar mass formed in a recent 20\,Myr-old starburst (\texttt{sfh.f\_burst}) are sampled within 50-500 Myr, 30-3000 Myr and 0.0-0.3, respectively, during the \texttt{CIGALE} modelling. The nebular emission is considered to be at solar metallicity. The dust attenuation is considered to follow a Milky Way-like attenuation curve from \citet{Calzetti+00}, with the equivalent E(B-V) ranging between 0 and 8 and the slope delta of the power law modifying the attenuation curve (\texttt{attenuation.powerlaw\_slope}) ranging from -0.5 to 0.5. For the dust emission, we adopt the dl2014 model from \citet{Draine+14}. The fraction of PAH emission (qpah) is fixed to 2.5 due to the lack of constraints by the shallow mid-IR data. The minimum starlight intensity (\texttt{dust.umin}), power-law exponent for the starlight intensity distribution (\texttt{dust.alpha}) and mass fraction of warm dust (\texttt{dust.gamma}) are sampled within 1-50, 1.0-3.0 and 0.0-1.0, respectively. The redshift is sampled between 3 and 9 with a fixed step of 0.1. 
The \texttt{CIGALE} modelling returns a $z_{\rm phot}=6.69^{+1.94}_{-1.62}$ for AC-2168, with the probability density of $z_{\rm phot}$ illustrated in Fig.~\ref{fig:redshift}. Although the sparse detections in the SED prevent a tight constraint on z$_{\rm phot}$, the non-detection in all but F444W, the IR-luminous nature and the red colour of the far-IR to millimetre SED \citep{Casey+18, Cooper+22} still largely rule out low- to intermediate-redshift solutions at z$<$4, while showing no specific preference to redshift solutions between 5 and 9. 

To further constrain the redshift of AC-2168, we search for far-IR emission lines in the spectrum of AC-2168 extracted at the continuum position in ALMA and NOEMA observations. By inspecting the 1.2\,mm spectrum from the two sidebands of the NOEMA observation, we clearly identify an emission peak at a central frequency of 249.1\, GHz with an integrated S/N=5.1, as shown in Fig.~\ref{fig:redshift}. We fit the line with a single Gaussian model and measure the line width of 520$\pm$110 km/s and total flux of 1.16$\pm$0.23 Jy km/s simultaneously. The central frequency of this line is consistent with the [\ion{C}{II}] 158$\mu$m emission at $z_{\rm spec}=6.631\pm0.001$, which falls in the centre of the broad plateau in $z_{\rm phot}$ probability distribution in Fig.~\ref{fig:redshift}. This redshift solution is also supported by the four mid-J CO lines (CO(6-5), CO(7-6), CO(9-8) and CO(10-9)) tentatively detected in Band 3 and Band 4 spectra. The spectra in these two bands are extracted from cleaned datacubes centred on the 2mm continuum. Given the matched beam size (see Sect.~\ref{sbsc:almaobs}), the same $r=0.6\arcsec$ aperture as NIRCam photometry is applied during the extraction in these two bands. Although none of these CO lines reach S/N to claim robust individual detection in the ALMA spectral scan, we perform an inverse variance weighted stacking of the spectra resampled to the same channel width of 100\,km/s and centred on the observed frequency of these four CO lines at $z_{\rm spec}=6.631$, which reveals an emission signal at $>$4 sigma significance (Fig.~\ref{fig:redshift}). The peak of the stacked signal is shifted by $\sim$100 km/s in velocity with respect to the centre of [\ion{C}{II}], while they are still consistent within the FWHM of these emission lines. As shown in Fig.~\ref{fig:redshift}, the spectra centred on each CO transition reveal tentative emission within the FWHM of the [\ion{C}{II}], suggesting the stacked signal in Band 3 and Band 4 is unlikely to be contributed by one or a few spurious spikes by chance. The detection of both [\ion{C}{II}] and CO emission, either individually or through stacking, supports the reliability of the $z_{\rm spec}=6.631$ solution. This $z_{\rm spec}$ confirmation also translates the measured continuum FWHM to a size of 1.1$\pm$0.5 kpc, which matches the DSFGs at a lower redshift of $z=3-6$ \citep{Ikarashi+15,Hodge+25}.

\subsection{SED modeling, Stellar mass and SFR}\label{sbsc:msfr_sed}

\begin{figure*}
    \centering
    \includegraphics[width=0.95\linewidth]{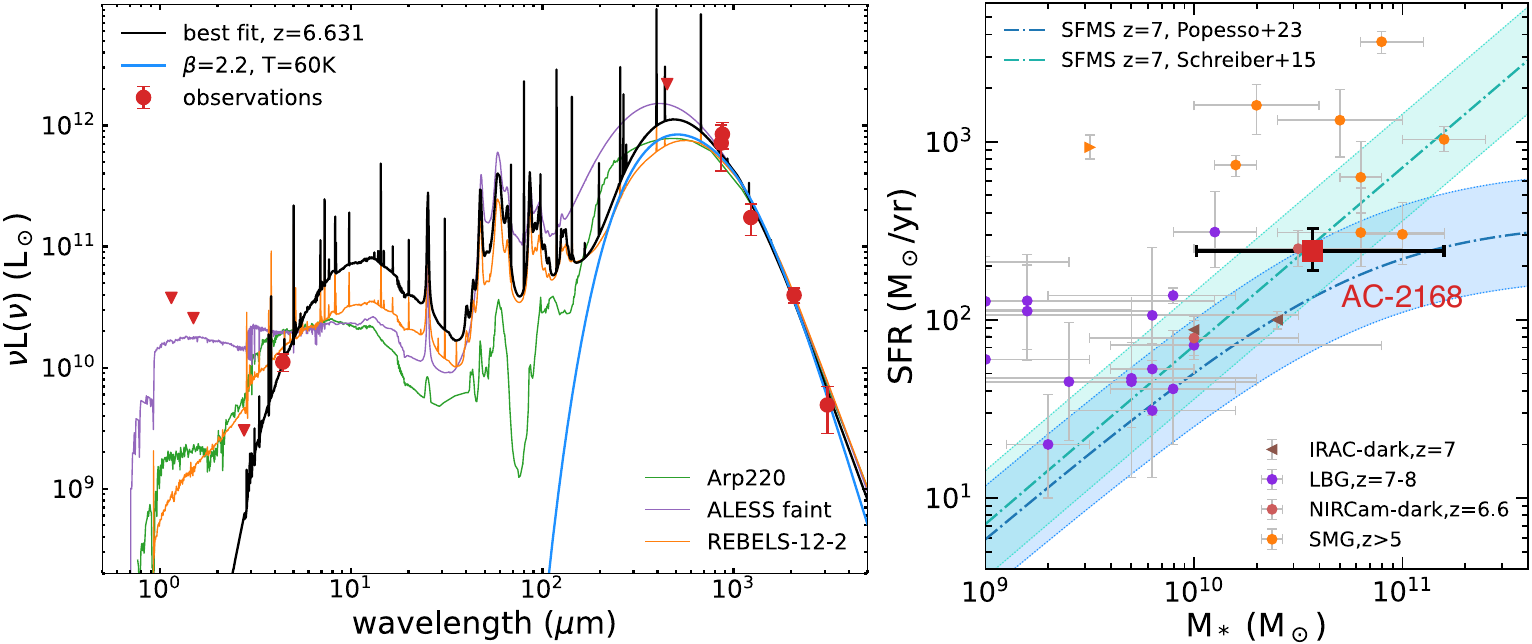}
    \caption{Left: the best-fit \texttt{CIGALE} full-wavelength model (black) and the best-fit modified blackbody model (blue) of AC-2168 SED compared to the SEDs of ULIRG Arp220, optically dark ALESS DSFGs and REBELS-12-2, an IRAC-dark DSFG in \citet{Fudamoto+21} at z=7.35. The flux and 2$\sigma$ flux upper limits of AC-2168 are marked in circles and triangles, respectively. The SEDs of literature sources are shifted to z=6.631 and normalised to AC-2168 at 1.2\,mm. Right: the best-fit SFR and M$_*$ of AC-2168 compared to literature samples of high-z DSFGs \citep{Capak+11,Riechers+13,Marrone+18,Williams+19,Casey+19,Jin+19,Riechers+20,Fudamoto+21,Sun+24,Xiao+24,Sun+25}, massive galaxies \citep{Dayal+22,Schouws+22}, and the star-formation main sequence at z=7 \citep{Schreiber+15,Popesso+23}.}
    \label{fig:sedfit}
\end{figure*}


We measure the stellar, star formation and dust properties of AC-2168 based on the multiwavelength photometry in Table~\ref{tab:photometry}. The modelling is performed with \texttt{CIGALE} using the same sampling of parameters as described in Sect.~\ref{sbsc:zspec} except for the redshift, which is fixed to the derived $z_{\rm spec}$. Physical properties derived from SED fitting are summarised in Table~\ref{tab:physical_properties}. Fig.~\ref{fig:sedfit} shows the best-fit SED models of AC-2168 with respect to the multiwavelength photometry. As a comparison, we also plot the SED templates of three galaxies: the prototype ULIRG Arp220 SED from \citet{Polletta+07}, optically faint ALESS DSFGs at z$\sim$3 \citep{daCunha+15} and a z$\sim$6.7 DSFG of lower M$_*$ that is only tentatively detected in deep IRAC observation \citep{Fudamoto+21}. All SED templates are normalised to 1.2 mm flux density of the \texttt{CIGALE} best-fit model of AC-2168. 

It is evident that AC-2168 is more obscured in the rest-frame optical than all of these template sources, even when compared to the IRAC-dark galaxy REBELS-12-2 \citep{Fudamoto+21} at z=7.35. Specifically, we highlight that AC-2168 is fainter in rest-frame optical bands than the most obscured subgroup of DSFGs detected by blind (sub) millimetre surveys \citep[i.e.][]{daCunha+15}. When shifted to $z=6.63$ and with flux normalised to AC-2168 at 1.2\,mm, all other DSFG templates present in Fig.~\ref{fig:sedfit} remain above the detection limit of COSMOS-Web and show 2-10 times higher fluxes in F277W and F444W compared to AC-2168. The extreme dust obscuration of AC-2168 is also reflected by the large dust attenuation derived from the SED fitting. As a comparison, typical high-z DSFGs selected by submillimeter observations and/or H-dropout/HST-dark nature in near IR have typical Av between 1.0 and 5.0 \citep{Wang+19,Barrufet+23a}. AC-2168 shows a large but not uniquely high attenuation of Av=5.4 magnitude in our SED modelling, indicating its faintness in the rest-frame optical is jointly contributed by the high redshift and the extreme obscuration compared to typical DSFGs. 

The multiwavelength continuum observations from ALMA and SCUBA2 provide good sampling on the far-IR SED of AC-2168, where the \texttt{CIGALE} fit returns an L$_{\rm IR}$ of $\rm 1.6^{+0.6}_{-0.4}\times10^{12}$ L$_\odot$. The corresponding SFR, as inferred from L$_{\rm IR}$ based on the calibration in \citep{Kennicutt+12}, reaches $244^{+82}_{-55}$ M$_\odot$/yr. In contrast, the stellar mass derived from \texttt{CIGALE} shows large uncertainty at $3.7^{+12.2}_{-2.6}\times10^{10}$ M$_\odot$, which is primarily due to the lack of detections and large flux uncertainties in the optical and infrared parts of the rest frame galaxy SED. The inferred M$_*$ and SFR put AC-2168 at the massive end of the star formation main sequence at z$\sim$6.6 \citep{Schreiber+15,Popesso+23}, as shown in Fig.~\ref{fig:redshift}. Remarkably, the moderate star formation activity is in contrast to the handful of bright SMGs known at z$_{\rm spec}>$5, which usually reach $\sim$1000 M$_{\odot}$/yr. This distinct characteristic, together with other properties discussed in Sect.~\ref{sbsc:mgas_tdep}, suggests that AC-2168 might be at a different evolutionary stage compared to most of the other high-z massive and/or dusty galaxies with active ongoing star formation (see Sect.~\ref{sbsc:qgprogenitor} for further discussion).

\begin{table}
	\centering
	\caption{Physical properties AC-2168}
	\label{tab:physical_properties}
	\begin{tabular}{ll} 
            \hline\hline
            z$_{\rm spec,[\ion{C}{II}]}$ & 6.631$\pm$0.001 \\
            S$_{\rm [\ion{C}{II}]}$ (Jy$\cdot$km/s) & 1.2$\pm$0.2 \\
            FWHM$_{\rm [\ion{C}{II}]}$ (km/s) & 520$\pm$110 \\
            L$_{\rm [\ion{C}{II}]}$ (L$_\odot$) & 1.3$\pm$0.3$\times$10$^{9}$\\
            FWHM$_{\rm cont}$ (kpc) & 1.1$\pm$0.5 \\
            \hline
		M$_*$ (M$_\odot$)& 3.7$^{+12.2}_{-2.6}\times10^{10}$  \\ 
		SFR (M$_\odot$/yr) & 244$^{+82}_{-55}$ \\
            L$_{\rm IR}$ (L$_\odot$) & 1.6$^{+0.6}_{-0.4}\times10^{12}$ \\
            A$_v$ (mag) & $5.4^{+1.6}_{-1.6}$ \\ 
            \hline
            M$_{\rm dust}$ (M$_\odot$) & 3.0$^{+0.8}_{-0.5}\times10^{8}$ \\ 
            T$_{\rm dust}$ (K) & 60$^{+11}_{-11}$ \\
            $\beta_{\rm dust}$ & 2.2$^{+0.6}_{-0.6}$ \\
		M$_{\rm gas}$ (M$_\odot$) & 4.1$\pm$0.8$\times$10$^{10}$ \\
		\hline
	\end{tabular}
\end{table}

\subsection{Gas mass and depletion time}\label{sbsc:mgas_tdep}

We probe the physical properties of multi-phase ISM with the multiwavelength photometry in far-IR to millimetre, as well as the far-IR emission lines covered by ALMA and NOEMA observations. With the millimetre continuum fluxes from SCUBA2, ALMA and NOEMA, we first derive the dust mass with the full-wavelength \texttt{CIGALE} modelling, as already described in Sect.~\ref{sbsc:zspec} and Sect.~\ref{sbsc:msfr_sed}. The fit results in a high dust mass (M$_{\rm dust}$) of $\rm 3.0^{+0.8}_{-0.5}\times10^8M_\odot$, reaching 0.8\% of the stellar mass of AC-2168. The high dust mass of AC-2168 matches the picture of highly dust-obscured star-forming galaxies, albeit at a high redshift compared to most known examples of this population. 

In addition to the dust mass, we also measure the dust temperature (T$_{\rm dust}$) by fitting a modified black body spectrum on the far-IR to millimetre photometry, as illustrated in Fig.~\ref{fig:redshift}. The heating and dimming of dust emission due to increased CMB temperature \citep{daCunha+13} is considered in the model, while we confirm that their impact on the inference of parameters is insignificant. Thanks to the deep submillimeter observation from STUDIES, we obtain a robust constraint on the T$_{\rm dust}$ at 60$\pm$11 K. Although initially selected at a relatively long wavelength millimetre observation, the derived T$_{\rm dust}$ of AC-2168 is not as biased low relative to the T$_{\rm dust}$-z relation \citep{Bethermin+15, Schreiber+18} as expected by simulations and seen in millimetre surveys \citep{Jin+22,Bing+24,Bethermin+25}. On the contrary, AC-2168 has an even higher T$_{\rm dust}$ than the average of UV-bright normal star-forming galaxies at similar redshift \citep[e.g.,][]{Dayal+22,Schouws+22}. This may indicate a geometry and/or star formation mode that differs from normal star-forming galaxies, which is discussed in more detail in Sect.~\ref{sbsc:qgprogenitor}. 

The detection of far-IR emission lines allows us to further estimate the total mass (M$_{\rm gas}$) and depletion time ($\tau_{\rm dep}$) of the gas reservoir of AC-2168. Although our data do not cover low-J CO lines commonly used to trace cold gas, [\ion{C}{II}] provide an alternative and robust probe to the molecular gas in galaxies, with a  [\ion{C}{II}]-to-H$_2$ conversion factor ($\alpha_{\rm [\ion{C}{II}]}$) stable across a wide range of redshifts and populations \citep{Zanella+18}. Using an $\alpha_{\rm [\ion{C}{II}]}=31$ from \citet{Zanella+18}, we derive the total cold gas mass in AC-2168 as 4.1$\pm$0.8$\times$10$^{10}$ M$_\odot$. Given the M$_{\rm dust}$, the derived M$_{\rm gas}$ results in a gas-to-dust ratio ($\delta_{\rm GDR}$) of $\sim$128. Metal-enriched massive galaxies from the local to the early Universe have been consistently found to exhibit gas-to-dust ratios in the range $ 100<\delta_{\rm GDR}<200$ \citep{Leroy+11,Berta+16, DeVis+21, Hagimoto+23}. Our measurement also falls within this regime, consistent with the picture of a metal-enriched, massive dusty star-forming galaxy. The derived M$_{\rm gas}$ led to a gas fraction ($f_{\rm gas}=M_{\rm gas}/(M_*+M_{\rm gas})$) of $\sim$52\% in AC-2168. This $f_{\rm gas}$ is slightly lower than massive SFGs at z$>$4 \citep{Riechers+10,Pavesi+19,Liu+19,Dessauges+20,Alvarez-Marquez+23} with $f_{\rm gas}\gtrsim0.7$. The joint effort of [\ion{C}{II}] and SED modelling also constrains the gas depletion time to $\sim$170 Myr. This is also consistent with the extrapolation of the redshift evolution of $\tau_{\rm dep}$ on the main sequence to the source redshift. However, compared to the REBELS sample \citep{Aravena+24}, the $\tau_{\rm dep}$ of AC-2168 is around a factor of 3 shorter, indicating a more rapid gas consumption by star formation in AC-2168 than normal but lower mass SFGs at a similar redshift.

\section{Anchoring NIRCam-Dark Galaxies in the Cosmic Evolution of Massive Galaxies}\label{sect:discussion}

\subsection{NIRCam-dark galaxies as the possible progenitor of earliest quiescent galaxies}\label{sbsc:qgprogenitor}

Massive DSFGs have been proposed to be the progenitors of QGs in the early Universe. Residing at $z_{\rm spec}$=6.631, AC-2168 adds to the handful of DSFGs spectroscopically confirmed at $5<z<7$ to date. The majority of these sources, especially those discovered by blind surveys, are confirmed to be extreme starbursts with SFR$>$1000 $M_\odot$/yr. AC-2168, in contrast, shows moderate SFR at its high M$_\star$, which puts it more likely within the star formation main sequence (SFMS) at z=6-7 (Fig.~\ref{fig:sedfit}). In addition to the normal SFR, the short $\tau_{\rm dep}$ relative to SFGs at similar redshift, the compact size in dust emission and the high T$_{\rm dust}$ are all consistent with the properties of massive compact SFGs prevalently found at z=2-4, which are proposed to be in the final stages on an evolutionary pathway to quiescence \citep[e.g.,][]{Dekel+09, Barro+13, Elbaz+18, Gomez-Guijarro+22b}. The f$_{\rm gas}$=52\% in AC-2168 is not as low as that in z=2-4 massive compact star-forming galaxies (0.1-0.4 in general, see \citet{Gomez-Guijarro+22b}). However, we caution that the large uncertainties due to the poorly constrained M$_{*}$ caused by the faintness of AC-2168 in NIRCam data mean that the f$_{\rm gas}$ of AC-2168 is still consistent with the low values in compact SFGs.  

\begin{figure*}
    \centering
    \includegraphics[width=0.95\linewidth]{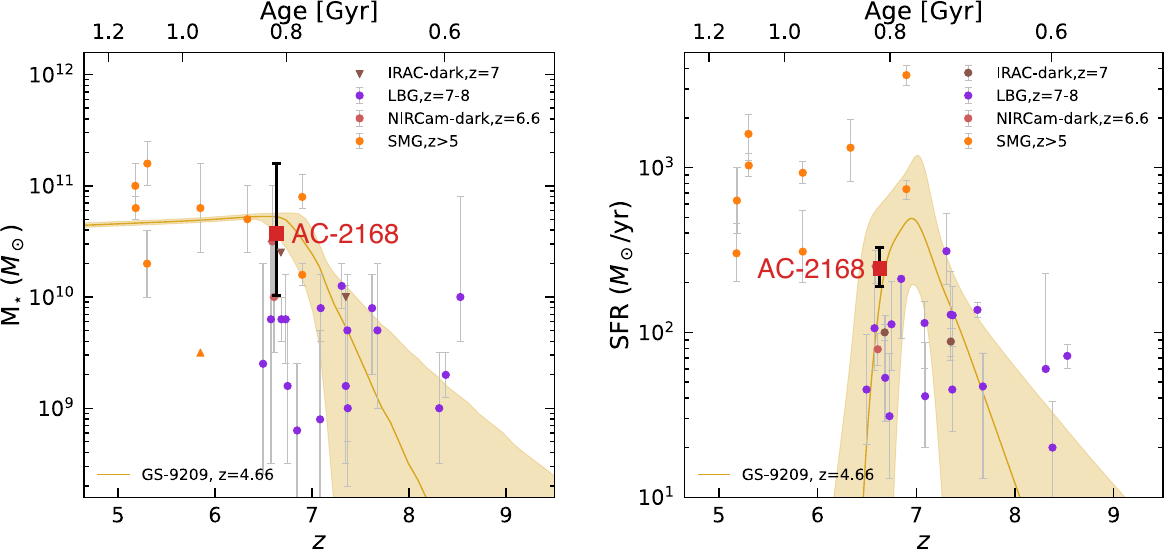}
    \caption{Comparison between the $M_*$ and SFR of AC-2168 and other massive star-forming/starburst galaxies in literature with the inferred mass assembly history (left) and SFH (right) of GS-9209, a prototype of the earliest massive QGs at z=4.66. The shaded regions mark the 1-$\sigma$ uncertainty of the mass assembly history and SFH of GS-9209 based on medium-resolution NIRSpec observations from \citet{Carnall+24}.}
    \label{fig:sfh}
\end{figure*}

The QG progenitor nature of AC-2168 is further evidenced by the proximity of its M*, SFR and redshift to the assembly history of the first generation of massive QGs at z$>$4. In Fig.~\ref{fig:sfh}, we overplot the M$_*$ and SFR of AC-2168 together with the mass assembly history and SFH of GS-9209, a representative massive QG at z=4.6 \citep{Carnall+23}. As inferred from the high S/N NIRSpec medium-resolution spectra, GS-9209 experienced rapid growth with a peak SFR of a few hundred M$_\odot$/yr within $\sim$200 Myrs at z$\sim$7. It is clearly shown that both the current mass and SFR of AC-2168 match within the 1-$\sigma$ uncertainties of the progenitor SFR and M$_{*}$ of GS-9209 at the peak of its SFH, followed by rapid star formation quenching. The short $\tau_{\rm dep}$ of AC-2168 is also comparable to the brief period of mass assembly of high-z QGs \citep[e.g.,][]{Valentino+23,Nanayakkara+24,Carnall+24}, further indicating the direct QG-progenitor nature of AC-2168. Other literature starbursts and SFGs seldom coincide with GS-9209-like QG progenitors at their peak of assembly simultaneously in $M_*$, SFR, redshift and formation timescale. As shown in Fig.~\ref{fig:sedfit}, blindly selected extreme starbursts in millimetre (see Fig.~\ref{fig:sedfit}) mostly have a factor of a few higher SFR at lower redshift, while the majority of massive SFGs at comparable or higher redshift better \citep[e.g.,][]{Dayal+22,Schouws+22} match with earlier epochs in the assembly history of high-z QGs. This evidence suggests that NIRCam-dark sources like AC-2168 might provide the best insight into the physical driver of the rapid formation and quenching of high-z QGs in its brief but vigorous growth in the first Gyr of the Universe. 

\begin{figure}
    \centering
    \includegraphics[width=1.0\linewidth]{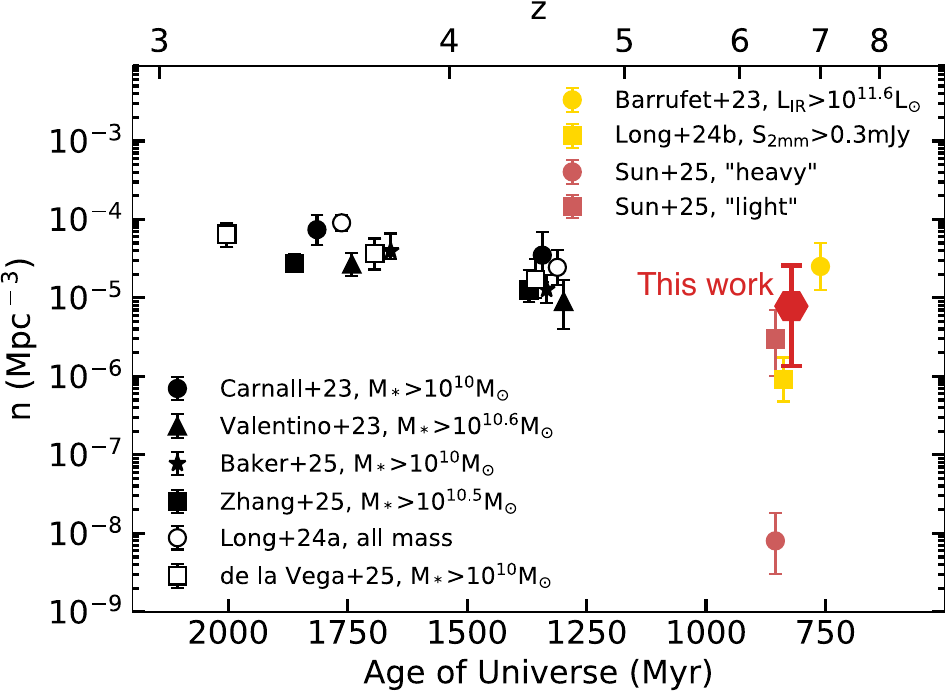}
    \caption{Our measurement of the number density of NIRCam-dark galaxies in comparison with the z=3-5 spectroscopically \citep{Carnall+23,Valentino+23,Baker+25,Zhang+25} or photometrically selected \citep{Long+24a,delaVega+25} massive QGs, z$\sim$6-7 dusty galaxies from REBELS \citep{Barrufet+23a} and exMORA \citep{Long+24b}, as well as the NIRCam-dark galaxies in ASPIRE \citep{Sun+25} under their "light" and "heavy" models.}
    \label{fig:density}
\end{figure}

The possible QG-progenitor nature of AC-2168 is also reflected by the consistency of the number density at z=6-7 of the source of its kind with massive QGs at z$\sim$4. As a serendipitously detected 2mm source, AC-2168 alone allows us to measure the number density of sources of its kind with a relatively simple estimation of the effective area and volume from the archival 2mm observations in A3COSMOS. The effective survey area ($\rm \Omega_{eff}$) for detecting AC-2168 in A3COSMOS 2mm data is derived from Eq.~\ref{eq:effarea}, following the method applied in ALPINE \citep{Bethermin+20} and N2CLS surveys \citep{Bing+23}.

\begin{equation}
    \Omega_{\rm eff} = \sum_{i}^{} \Omega(\rm \sigma_i)\mathcal{C}(\sigma_i) 
    \label{eq:effarea}
\end{equation}

$\Omega(\sigma_i)$ and $\mathcal{C}$($\sigma_i$) in the equation represents the A3COSMOS coverage with a noise level of $\sigma_i$ in 2\,mm continuum and the expected completeness of point-like sources with the 2\,mm flux density of AC-2168 within this coverage. The completeness of detection under a certain noise level is set to follow the Gaussian theorem, which has been proved to be a good approximation in deep continuum observations \citep{Bethermin+20,Bing+23}. Given the observed z$_{\rm spec}$ of AC-2168, we derive the effective co-moving volume of the A3COSMOS 2\,mm continuum observation on AC-2168 using a redshift range of z=6-7. As the only source confirmed in the volume without robust detection in the COSMOS-Web NIRCam source catalogue, it turns to a number density of $7.8^{+18.0}_{-6.5}\times10^{-6}$ cMpc$^{-3}$, where the uncertainty is dominated by Poisson noise at the small number of our sample. 

We compare the derived number density of NIRCam-dark galaxies with those of the first generation of massive QGs at z=3-5 \citep[e.g.,][]{Carnall+23,Valentino+23,Long+24a,delaVega+25, Zhang+25}. As illustrated in Fig.~\ref{fig:density}, if we take the median of the number density of z=4-5 QGs in literature at 1.8$\times10^{-5}$ cMpc$^{-3}$, the number density of NIRCam-dark sources from our analysis reaches $\sim$42\% of that of z=4-5 QGs, and the fraction is even fully consistent with 100\% within the uncertainties. This orders-of-magnitude consistency in abundance indicates that NIRCam-dark galaxies like AC-2168 might represent the progenitor of a significant fraction of the most massive QGs formed in the first 1.5 Gyr of the Universe. The redshifts of the two populations translate to a $\sim$510 Myr difference in the age of the universe. Given the short gas depletion time of 170 Myr (see Sect.~\ref{sbsc:mgas_tdep}) for AC-2168, the shutdown of star formation, even if no gas consumption mechanism other than star formation is accounted for, could be completed well before the time when similarly massive QGs like RUBIES-EGS-QG-1 \citep{deGraaff+25} and GS-9209 \citep{Carnall+23} were observed. The comparable abundance relative to QGs and short $\tau_{\rm dep}$, in addition to the match in mass assembly history, jointly strengthens that NIRCam-dark galaxies represent the progenitor of at least some of the most massive QGs formed in the first 1.5 Gyr of the universe. 

Apart from the high-z quiescent galaxies, we further compare the number density of our measurements with the studies on dusty galaxies at z=6-7 \citep{Barrufet+23b,Long+24b} and similar NIRCam-dark populations in the proximity of two z$\sim$6.5 quasars in the ASPIRE sample \citep{Sun+25}. The number density of NIRCam-dark sources from our measurements reaches $\rm \sim31\%$ of the normal high-z star-forming galaxies in REBELS with ULIRG-like IR luminosities, and the two number densities are fully consistent within the 1$\sigma$ uncertainties. The ex-MORA sample, although still consistent with our sample within 1$\sigma$, drops almost an order of magnitude lower in its number density. This difference is not surprising if we take the shallow survey depth of ex-MORA into account. With continuum RMS of $\rm \sim90\mu Jy/beam$ at 2\,mm, ex-MORA is unlikely to detect other AC-2168-like DSFGs with moderate star formation but dominantly probe the more IR luminous DSFG population of lower number density. The comparison with the NIRCam-dark sample in \citep{Sun+25} under two different scenarios shows a clear preference for the "light" model, which assumes uniform occupation of NIRCam-dark sources in dark matter halos with mass above $10^{11.4}$ M$_{\odot}$. The "heavy" model, assuming a strong bias to the most massive quasar-hosting halos, however, leads to a number density of 3 orders of magnitude lower than our estimation. These results further imply the remarkable contribution of the NIRCam-dark galaxy to the bulk of z$>$6 DSFGs, which are believed to be the progenitors of high-z QGs.

\subsection{Diverse environment of NIRCam-dark galaxies}\label{sbsc:environment}


A decade ago, numerical simulations and statistical analysis on survey data suggested the close connection between the formation and growth of galaxies and dense environments, such as protoclusters, in the early Universe \citep{Chiang+13, Bethermin+13}. More recently, the widespread association between rapidly assembled massive galaxies and dense environments at z$\sim$4 has been further revealed by various observations \citep[e.g.,][]{Calvi+21, Brinch+23,Tanaka+24, Kakimoto+24, deGraaff+25}. This supports the expectation of galaxy formation models and cosmological simulations, which predict the dominant contribution of high-$z$ overdensities in the star formation in the early Universe \citep{Chiang+17,Shimakawa+18} fed by cold gas streams in hot media \citep{Dekel+13}. The association with galaxy overdensities is also found to be common among DSFGs at the cosmic dawn \citep[e.g.,][]{Miller+18, Casey+19,WangGeorge+21,Sun+24,Herard-Demanche+25,Akins+25,Lagache+25}. These include the recently identified highly obscured NIRCam-dark galaxies at a similar redshift of AC-2168 \citep{Sun+25}. Their samples are detected in the vicinity of luminous quasars, most of which are known to associate with massive dark-matter halos \citep{Pizzati+24} likely hosting protoclusters \citep{Kashino+23,Wang+23,Champagne+25,Champagne+25b}. 


\begin{figure*}
    \centering
    \includegraphics[width=0.95\linewidth]{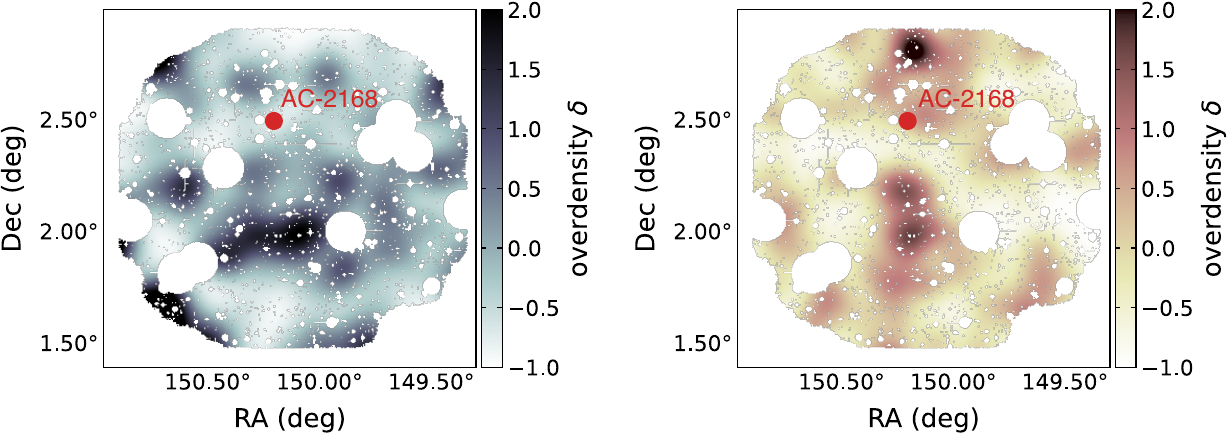}
    \caption{The position of AC-2168 overlaid on the galaxy overdensity maps at z$\sim$6.6 traced by LAE (left) and LBGs (right) from the CHORUS survey \citep{Inoue+20,Yoshioka+22}.}
    \label{fig:overdensity}
\end{figure*}

As a galaxy of a similar kind, we explore the local environment of AC-2168 using LAEs and LBGs at z$=$6.4-7.0. LAEs and LBGs are widely used to probe high-z galaxy overdensities for their simple and efficient colour or dropout selection with combinations of wideband and medium/narrow band imaging. Previous studies have used LAEs/LBGs to reveal the overdensity around massive quiescent galaxies \citep[e.g.,][]{Tanaka+24, Kakimoto+24}, dusty galaxies \citep[e.g.,][]{Casey+19, Sun+24}) and quasars \citep[e.g.,][]{Kashino+23,Wang+23,Champagne+25,Champagne+25b} at $z>4$, suggesting the consistency of LAEs/LBGs in probing the environment of different types of massive galaxies in the early Universe.

The LAEs and LBGs used to trace galaxy overdensities in the COSMOS field are identified by the CHORUS survey \citep{Inoue+20, Yoshioka+22} using Subaru HSC NB921 and IB945 observations in the COSMOS field down to 26.3 and 25.9 mag, respectively. Combining with broadband imaging data in the same field, the narrow- and medium-band Subaru HSC observation with the NB921 filter from CHORUS provides efficient selection of LAEs at $z=6.54-6.63$, which roughly covers the $z_{\rm spec}$ of AC-2168. As a comparison, the observations with IB945 are effective to probe LBGs at $z=6.4-6.9$, providing wider redshift coverage that fully encloses the possible redshift distribution of proto-cluster members. More detailed information on LAE and LBG selection at the redshift range of our interest can be found in \citet{Yoshioka+22}. 

The production of overdensity maps with LAEs and LBGs is described in detail in \citet{Yoshioka+22} as well. In Fig.~\ref{fig:overdensity}, we compare the location of AC-2168 with LAE/LBG densities in the COSMOS field. 
Although the majority of known massive DSFGs at $z>5$ tend to be associated with overdensities, we find that AC-2168 lacks coincidence with the density peaks traced by LAE/LBGs at similar redshift. As a comparison, the two NIRCam-dark sources at similarly high redshift in \citet{Sun+25} are both discovered in the vicinity of luminous quasars, showing clear association with high-z galaxy overdensities confirmed by narrow band or spectroscopic observation \citep[e.g.,][]{Wang+23,Champagne+25,Champagne+25b}. 

Despite the preference for emergence in protoclusters, studies also indicate that the formation of massive galaxies in the early Universe does not always require a high-density environment. Hydrodynamical simulations like FLARES reveal that overdensities enhance the abundance of massive galaxies but do not monopolise them \citep{Lovell+21}. It finds a non-negligible number of massive ($\rm M_*>10^{10} M_\odot$) galaxies at $z>5$ in regions with local density as low as the cosmic average. Various hydro simulation frameworks also reveal the critical roles of anisotropic tidal fields/cosmic-web geometry in enhancing the gas inflow to galaxies in the early Universe \citep{DiMatteo+17,Liao+19,Song+20}, indicating the rapid assembly of massive galaxies might not strictly require an overdense environment. From the perspective of observations, studies also find examples indicating that z$>$3 massive quiescent galaxies may complete the assembly of the bulk of their stellar mass outside of the dense core of high-z (proto)clusters before finally falling into the cores \citep{Jin+24}. The lack of coincidence of AC-2168 with overdensities, in this case, probably provides live evidence of the rapid formation and quenching of massive galaxies in the outskirts or outside of overdensities. 

Notably, the lack of physical association with galaxy overdensities also provides self-consistent validation on the consistent number density from our estimation with those derived under "light model" in \citep{Sun+25} (Fig.~\ref{fig:density}). Comparing to the "heavy model" populating NIRCam-dark galaxies in only the most massive halos hosting quasars, the "light model" assumes constant occupation fraction of NIRCam-dark galaxies in dark matter halos down to log(M$_{\rm halo}$/M$_\odot$)$=$11.4, which falls an order of magnitude lower than the halo mass of LAE/LBG selected protoclusters at z=6-7 in COSMOS \citep{Brinch+24}. As a serendipitous detection in ALMA data with redshift uncorrelated to the source at the phase centre, our selection of AC-2168 is less biased towards massive halos than \citet{Sun+25}. With these prerequisites in mind, the consistency between our results and the "light model" number density suggests that NIRCam-dark galaxies are not uniquely located in the most massive halos, which is consistent with the lack of spatial coincidence between AC-2168 and LAE/LBG protoclusters.  

Our observation, combined with \citet{Sun+25}, indicates a substantial diversity of the environment where the most obscured early massive galaxies form and grow. However, we are also aware that the evidence of the mismatch with overdensities relies on identification based on LBG and LAE selection, which is only sensitive to UV bright sources with low dust attenuation. Star-forming galaxies at comparable or higher redshift could already be mature enough to produce a significant amount of dust, making a substantial fraction of them faint or even completely invisible in rest-frame UV \citep[e.g.,][]{Fudamoto+21,Dayal+22,Schouws+22,Barrufet+23b}. Although the cosmic evolution of the density of dusty galaxies is still not well constrained, they are known to dominate the massive end of the star-forming population up to at least $z\sim4-5$. Overdensity cores dominated by these dusty sources at high redshift have also been identified in previous studies\citep[e.g.,][]{Miller+18, WangGeorge+21}. Thus, we could not exclude that an overdensity abundant in dusty sources around AC-2168 is missed by the LAE/LBG selection.

\section{Conclusions}

We report the discovery of AC-2168, a NIRCam-dark, millimetre-bright galaxy at $z_{\rm spec}=6.631$ serendipitously detected in ALMA 2mm continuum and spectroscopically confirmed with NOEMA 1.2mm, ALMA 2mm and ALMA 3mm observations. The key results are summarised as follows:

\begin{enumerate}
    \item We confirm the spectroscopic redshift of this galaxy at $z_{\rm spec}=6.631$ through a robust detection of \ion{[C}{ii]}~158\,$\mu$m and tentative detections of four CO lines. Despite deep JWST imaging, the source remains undetected shortward of 4.4\,$\mu$m and only tentatively detected in F444W data, which is in contrast to the majority of high-z DSFGs in previous studies. 
    \item We perform SED modeling on AC-2168 with \texttt{CIGALE}, implying a moderate SFR of $\approx 2.4\times10^{2}\,M_\odot\,{\rm yr}^{-1}$ at likely high M$_*$ of $\approx 3.7\times10^{10} M_\odot$.  This places AC-2168 to be less active in star formation than the hyper-luminous ($\sim10^3\,M_\odot\,{\rm yr}^{-1}$) DSFGs commonly seen in wide area blind surveys at $z>6$, yet still firmly consistent with a galaxy undergoing vigorous, obscured growth at the high mass end of the star formation main sequence at its redshift.
    \item With the modelling on the IR SED and 2mm continuum data of AC-2168, we reveal warm ($T_{\rm dust} \sim 60$\,K) and compact (FWHM $\sim0.21"$ or $\sim$1.1 kpc at $z=6.631$) dust emission. Using \ion{[C}{ii]} as a gas mass tracer, we infer $M_{\rm gas} \sim 4\times10^{10}\,M_\odot$, $f_{\rm gas} \sim 52$\%, and $\tau_{\rm dep} \sim 170$\,Myr. These values are consistent with a rapid and efficient mass assembly episode capable of producing compact, massive descendants on short timescales.
    \item We dissect the evolutionary connection between NIRCam-dark galaxies at $z=6-7$ with the first generation of massive QGs at $z\gtrsim4$. The moderate SFR, short $\tau_{\rm dep}$, and high obscuration of NIRCam-dark galaxies offer a better match to the reconstructed SFHs of massive QGs at $z\gtrsim4$ than do hyper-luminous DSFGs. As AC-2168 is serendipitously detected, we also derive the density of NIRCam-dark galaxies at $z\simeq6$-7 at $n=7.8^{+18.0}_{-6.5}\times10^{-6}\,{\rm cMpc}^{-3}$, which also aligns with massive QGs at $z\gtrsim4$. Both evidence indicate that NIRCam-dark galaxies might representatively trace the short but vigorous assembly of z$\gtrsim$4 massive QGs.
    \item Using the UV-bright LAEs/LBGs selected from narrow/medium band imaging from Subaru HSC imaging, we find no overdensity at $z\sim6.5$-6.9 that is clearly associated with AC-2168. This is in sharp contrast to the two NIRCam-dark galaxies in \citep{Sun+25} discovered in the vicinity of luminous quasars in overdensities at similar redshift, implying the possible diverse environment hosting massive galaxies in the first Gyr of the Universe. 
\end{enumerate}

\noindent Our results argue that the rapid growth of a substantial fraction of massive galaxies at the end of the reionisation era might remain elusive even in the current JWST NIRCam surveys. Sensitive wide area, wide band observations in (sub-) millimetre, such as the ALMA after the Wideband Sensitivity Upgrade (WSU) \citep{Carpenter+23}, and future single-dish telescopes like the Atacama Large Aperture Submillimeter Telescope (AtLAST) \citep{Klaassen+19} and Xue-shan-mu-chang Submillimeter Telescope (XSMT) \citep{XSMT+25}, will play critical roles in refining their space density and luminosity function. On the level of individual sources, observations on low-$J$ CO, higher-resolution \ion{[C}{ii]} and continuum, and deep JWST rest-frame optical-IR imaging and spectroscopy will jointly strengthen the constraints on gas excitation, kinematics, sizes and stellar masses to test their connection to the first generation of massive QGs with details from more extended perspectives. JWST NIRCam WFSS observations over the surroundings of blindly detected NIRCam-dark sources will provide environmental demographics to further evaluate if and how tightly their emergence is associated with overdensities in the early Universe.

\section*{Acknowledgements}

We are grateful to Adam Carnall, Yoshinobu Fudamoto and Takehiro Yoshioka for kindly sharing the data products in their research. LB and SO acknowledge support from the Sussex Astronomy Centre STFC Consolidated Grant 2023-2026 (ST/X001040/1). MX acknowledges support from the Swiss State Secretariat for Education, Research and Innovation (SERI) under contract number MB22.00072, as well as from the Swiss National Science Foundation (SNSF) through project grant 200020\_207349. SA gratefully acknowledges the Collaborative Research Center 1601 (SFB 1601 sub-project C2) funded by the Deutsche Forschungsgemeinschaft (DFG, German Research Foundation) – 500700252. DL acknowledges the support from the Strategic Priority Research Program of the Chinese Academy of Sciences, grant No. XDB0800401. MF has received funding from the European Union’s Horizon 2020 research and innovation program under the Marie Sklodowska-Curie grant agreement No 101148925.
O.R.C. is supported by a National Science Foundation Astronomy and Astrophysics Postdoctoral Fellowship under award AST-2503202. SG acknowledges financial support from the Cosmic Dawn Center (DAWN), funded by the Danish National Research Foundation (DNRF) under grant No. 140. Support for this work was provided by NASA through grant JWST-GO-01727 awarded by the Space Telescope Science Institute, which is operated by the Association of Universities for Research in Astronomy, Inc., under NASA contract NAS 5-26555. This work is based on observations carried out under project number E24AI with the IRAM NOEMA Interferometer. IRAM is supported by INSU/CNRS (France), MPG (Germany) and IGN (Spain). This paper makes use of the following ALMA data: ADS/JAO.ALMA\#2013.1.00884.S, ADS/JAO.ALMA\#2019.1.01722.S. ALMA is a partnership of ESO (representing its member states), NSF (USA) and NINS (Japan), together with NRC (Canada), NSTC and ASIAA (Taiwan), and KASI (Republic of Korea), in cooperation with the Republic of Chile. The Joint ALMA Observatory is operated by ESO, AUI/NRAO and NAOJ. This research made use of APLpy, an open-source plotting package for Python hosted at http://aplpy.github.com. This research made use of Astropy, a community-developed core Python package for Astronomy \citep{astropy+13}.


The author contributions, following the CRediT taxonomy were {\bf Bing:} Conceptualization, Data curation, Investigation, Visualization, Writing -- original draft; {\bf Oliver:} Funding acquisition, Supervision, Writing – review \& editing; {\bf Xiao and Lagache:} Conceptualization, Writing – review \& editing; {\bf Adscheid, Liu, Magnelli and Neri:} Data curation, Writing – review \& editing; {\bf All other authors:} Writing – review \& editing.

\section*{Data Availability}

The ALMA data and flux measurements used in this paper are publicly available from the ALMA Science Archive (https://almascience.eso.org/aq/) and the A3COSMOS data page (https://sites.google.com/view/a3cosmos/data?authuser=0). NOEMA continuum and spectra are available upon request to the corresponding author. The JWST NIRCam images are available from the webpage of COSMOS-Web Public Data Release 1 (https://cosmos2025.iap.fr/).



\bibliographystyle{mnras}
\bibliography{example} 

@ARTICLE{astropy+13,
       author = {{Astropy Collaboration} and {Robitaille}, Thomas P. and {Tollerud}, Erik J. and {Greenfield}, Perry and {Droettboom}, Michael and {Bray}, Erik and {Aldcroft}, Tom and {Davis}, Matt and {Ginsburg}, Adam and {Price-Whelan}, Adrian M. and {Kerzendorf}, Wolfgang E. and {Conley}, Alexander and {Crighton}, Neil and {Barbary}, Kyle and {Muna}, Demitri and {Ferguson}, Henry and {Grollier}, Fr{\'e}d{\'e}ric and {Parikh}, Madhura M. and {Nair}, Prasanth H. and {Unther}, Hans M. and {Deil}, Christoph and {Woillez}, Julien and {Conseil}, Simon and {Kramer}, Roban and {Turner}, James E.~H. and {Singer}, Leo and {Fox}, Ryan and {Weaver}, Benjamin A. and {Zabalza}, Victor and {Edwards}, Zachary I. and {Azalee Bostroem}, K. and {Burke}, D.~J. and {Casey}, Andrew R. and {Crawford}, Steven M. and {Dencheva}, Nadia and {Ely}, Justin and {Jenness}, Tim and {Labrie}, Kathleen and {Lim}, Pey Lian and {Pierfederici}, Francesco and {Pontzen}, Andrew and {Ptak}, Andy and {Refsdal}, Brian and {Servillat}, Mathieu and {Streicher}, Ole},
        title = "{Astropy: A community Python package for astronomy}",
      journal = {\aap},
         year = 2013,
        month = oct,
       volume = {558},
          eid = {A33},
        pages = {A33},
          doi = {10.1051/0004-6361/201322068},
archivePrefix = {arXiv},
       eprint = {1307.6212},
 primaryClass = {astro-ph.IM},
       adsurl = {https://ui.adsabs.harvard.edu/abs/2013A&A...558A..33A}
}

@ARTICLE{Cooper+22,
       author = {{Cooper}, Olivia R. and {Casey}, Caitlin M. and {Zavala}, Jorge A. and {Champagne}, Jaclyn B. and {da Cunha}, Elisabete and {Long}, Arianna S. and {Spilker}, Justin S. and {Staguhn}, Johannes},
        title = "{Searching Far and Long. I. Pilot ALMA 2 mm Follow-up of Bright Dusty Galaxies as a Redshift Filter}",
      journal = {\apj},
         year = 2022,
        month = may,
       volume = {930},
       number = {1},
          eid = {32},
        pages = {32},
          doi = {10.3847/1538-4357/ac616d},
archivePrefix = {arXiv},
       eprint = {2203.14973},
 primaryClass = {astro-ph.GA},
       adsurl = {https://ui.adsabs.harvard.edu/abs/2022ApJ...930...32C}
}

@ARTICLE{Casey+18,
       author = {{Casey}, Caitlin M. and {Zavala}, Jorge A. and {Spilker}, Justin and {da Cunha}, Elisabete and {Hodge}, Jacqueline and {Hung}, Chao-Ling and {Staguhn}, Johannes and {Finkelstein}, Steven L. and {Drew}, Patrick},
        title = "{The Brightest Galaxies in the Dark Ages: Galaxies{\textquoteright} Dust Continuum Emission during the Reionization Era}",
      journal = {\apj},
         year = 2018,
        month = jul,
       volume = {862},
       number = {1},
          eid = {77},
        pages = {77},
          doi = {10.3847/1538-4357/aac82d},
archivePrefix = {arXiv},
       eprint = {1805.10301},
 primaryClass = {astro-ph.GA},
       adsurl = {https://ui.adsabs.harvard.edu/abs/2018ApJ...862...77C}
}

@ARTICLE{McKinney+23,
       author = {{McKinney}, Jed and {Manning}, Sinclaire M. and {Cooper}, Olivia R. and {Long}, Arianna S. and {Akins}, Hollis and {Casey}, Caitlin M. and {Faisst}, Andreas L. and {Franco}, Maximilien and {Hayward}, Christopher C. and {Lambrides}, Erini and {Magdis}, Georgios and {Whitaker}, Katherine E. and {Yun}, Min and {Champagne}, Jaclyn B. and {Drakos}, Nicole E. and {Gentile}, Fabrizio and {Gillman}, Steven and {Gozaliasl}, Ghassem and {Ilbert}, Olivier and {Jin}, Shuowen and {Koekemoer}, Anton M. and {Kokorev}, Vasily and {Liu}, Daizhong and {Rich}, R. Michael and {Robertson}, Brant E. and {Valentino}, Francesco and {Weaver}, John R. and {Zavala}, Jorge A. and {Allen}, Natalie and {Kartaltepe}, Jeyhan S. and {McCracken}, Henry Joy and {Paquereau}, Louise and {Rhodes}, Jason and {Shuntov}, Marko and {Toft}, Sune},
        title = "{A Near-infrared-faint, Far-infrared-luminous Dusty Galaxy at z {\ensuremath{\sim}} 5 in COSMOS-Web}",
      journal = {\apj},
         year = 2023,
        month = oct,
       volume = {956},
       number = {2},
          eid = {72},
        pages = {72},
          doi = {10.3847/1538-4357/acf614},
archivePrefix = {arXiv},
       eprint = {2304.07316},
 primaryClass = {astro-ph.GA},
       adsurl = {https://ui.adsabs.harvard.edu/abs/2023ApJ...956...72M}
}

@ARTICLE{Manning+25,
       author = {{Manning}, Sinclaire M. and {McKinney}, Jed and {Whitaker}, Katherine E. and {Long}, Arianna S. and {Cooper}, Olivia R. and {Casey}, Caitlin M. and {Arango-Toro}, Rafael C. and {Champagne}, Jaclyn B. and {Drakos}, Nicole E. and {Faisst}, Andreas L. and {Franco}, Maximilien and {Gozaliasl}, Ghassem and {Harish}, Santosh and {Hatamnia}, Hossein and {Hayward}, Christopher C. and {Hirschmann}, Michaela and {Kartaltepe}, Jeyhan S. and {Koekemoer}, Anton M. and {Liu}, Daizhong and {Magdis}, Georgios E. and {McCracken}, Henry Joy and {Rhodes}, Jason and {Robertson}, Brant E. and {Talia}, Margherita and {Valentino}, Francesco and {Weaver}, John R. and {Zavala}, Jorge A.},
        title = "{SCUBADive II: Searching for $z>4$ Dust-Obscured Galaxies via F150W-Dropouts in COSMOS-Web}",
      journal = {arXiv e-prints},
         year = 2025,
        month = may,
          eid = {arXiv:2505.09703},
        pages = {arXiv:2505.09703},
          doi = {10.48550/arXiv.2505.09703},
archivePrefix = {arXiv},
       eprint = {2505.09703},
 primaryClass = {astro-ph.GA},
       adsurl = {https://ui.adsabs.harvard.edu/abs/2025arXiv250509703M}
}

@ARTICLE{XSMT+25,
       author = {{XSMT Project Collaboration Group} and {Ao}, Yiping and {Chang}, Jin and {Chen}, Zhiwei and {Cui}, Xiangqun and {Du}, Kaiyi and {Du}, Fujun and {Gong}, Yan and {Han}, Zhanwen and {Herczeg}, Gregory and {Ho}, Luis C. and {Hu}, Jie and {Jing}, Yipeng and {Jiao}, Sihan and {Ju}, Binggang and {Li}, Jing and {Li}, Xiaohu and {Li}, Xiangdong and {Lin}, Lingrui and {Lin}, Zhenhui and {Liu}, Daizhong and {Liu}, Dong and {Liu}, Guoxi and {Lou}, Zheng and {Lu}, Dengrong and {Mao}, Ruiqing and {Miao}, Wei and {Qian}, Yuan and {Qiu}, Keping and {Shen}, Zhiqiang and {Shi}, Yong and {Shi}, Shengcai and {Shu}, Chenggang and {Sun}, Jixian and {Sun}, Xiaohui and {Sun}, Yichen and {Wang}, Junzhi and {Wang}, Ke and {Wang}, Na and {Wang}, Ran and {Wang}, Tao and {Wu}, Jingwen and {Wu}, Xiangping and {Wu}, Xuefeng and {Xiao}, Di and {Yao}, Qijun and {Yao}, Yong and {Zhang}, Wen and {Zhang}, Xuguo and {Zhang}, Zhiyu and {Zheng}, Yuanpeng},
        title = "{Scientific Objectives of the Xue-shan-mu-chang 15-meter Submillimeter Telescope}",
      journal = {arXiv e-prints},
         year = 2025,
        month = sep,
          eid = {arXiv:2509.13983},
        pages = {arXiv:2509.13983},
          doi = {10.48550/arXiv.2509.13983},
archivePrefix = {arXiv},
       eprint = {2509.13983},
 primaryClass = {astro-ph.GA},
       adsurl = {https://ui.adsabs.harvard.edu/abs/2025arXiv250913983X}
}

@ARTICLE{Riechers+10,
       author = {{Riechers}, Dominik A. and {Capak}, Peter L. and {Carilli}, Christopher L. and {Cox}, Pierre and {Neri}, Roberto and {Scoville}, Nicholas Z. and {Schinnerer}, Eva and {Bertoldi}, Frank and {Yan}, Lin},
        title = "{A Massive Molecular Gas Reservoir in the z = 5.3 Submillimeter Galaxy AzTEC-3}",
      journal = {\apjl},
         year = 2010,
        month = sep,
       volume = {720},
       number = {2},
        pages = {L131-L136},
          doi = {10.1088/2041-8205/720/2/L131},
archivePrefix = {arXiv},
       eprint = {1008.0389},
 primaryClass = {astro-ph.CO},
       adsurl = {https://ui.adsabs.harvard.edu/abs/2010ApJ...720L.131R}
}

@ARTICLE{Pavesi+19,
       author = {{Pavesi}, Riccardo and {Riechers}, Dominik A. and {Faisst}, Andreas L. and {Stacey}, Gordon J. and {Capak}, Peter L.},
        title = "{Low Star Formation Efficiency in Typical Galaxies at z = 5-6}",
      journal = {\apj},
         year = 2019,
        month = sep,
       volume = {882},
       number = {2},
          eid = {168},
        pages = {168},
          doi = {10.3847/1538-4357/ab3a46},
archivePrefix = {arXiv},
       eprint = {1812.00006},
 primaryClass = {astro-ph.GA},
       adsurl = {https://ui.adsabs.harvard.edu/abs/2019ApJ...882..168P}
}

@ARTICLE{Dessauges+20,
       author = {{Dessauges-Zavadsky}, M. and {Ginolfi}, M. and {Pozzi}, F. and {B{\'e}thermin}, M. and {Le F{\`e}vre}, O. and {Fujimoto}, S. and {Silverman}, J.~D. and {Jones}, G.~C. and {Vallini}, L. and {Schaerer}, D. and {Faisst}, A.~L. and {Khusanova}, Y. and {Fudamoto}, Y. and {Cassata}, P. and {Loiacono}, F. and {Capak}, P.~L. and {Yan}, L. and {Amorin}, R. and {Bardelli}, S. and {Boquien}, M. and {Cimatti}, A. and {Gruppioni}, C. and {Hathi}, N.~P. and {Ibar}, E. and {Koekemoer}, A.~M. and {Lemaux}, B.~C. and {Narayanan}, D. and {Oesch}, P.~A. and {Rodighiero}, G. and {Romano}, M. and {Talia}, M. and {Toft}, S. and {Vergani}, D. and {Zamorani}, G. and {Zucca}, E.},
        title = "{The ALPINE-ALMA [C II] survey. Molecular gas budget in the early Universe as traced by [C II]}",
      journal = {\aap},
         year = 2020,
        month = nov,
       volume = {643},
          eid = {A5},
        pages = {A5},
          doi = {10.1051/0004-6361/202038231},
archivePrefix = {arXiv},
       eprint = {2004.10771},
 primaryClass = {astro-ph.GA},
       adsurl = {https://ui.adsabs.harvard.edu/abs/2020A&A...643A...5D}
}

@ARTICLE{Yang+23,
       author = {{Yang}, Jinyi and {Wang}, Feige and {Fan}, Xiaohui and {Hennawi}, Joseph F. and {Barth}, Aaron J. and {Ba{\~n}ados}, Eduardo and {Sun}, Fengwu and {Liu}, Weizhe and {Cai}, Zheng and {Jiang}, Linhua and {Li}, Zihao and {Onoue}, Masafusa and {Schindler}, Jan-Torge and {Shen}, Yue and {Wu}, Yunjing and {Bhowmick}, Aklant K. and {Bieri}, Rebekka and {Blecha}, Laura and {Bosman}, Sarah and {Champagne}, Jaclyn B. and {Colina}, Luis and {Connor}, Thomas and {Costa}, Tiago and {Davies}, Frederick B. and {Decarli}, Roberto and {De Rosa}, Gisella and {Drake}, Alyssa B. and {Egami}, Eiichi and {Eilers}, Anna-Christina and {Evans}, Analis E. and {Farina}, Emanuele Paolo and {Habouzit}, Melanie and {Haiman}, Zoltan and {Jin}, Xiangyu and {Jun}, Hyunsung D. and {Kakiichi}, Koki and {Khusanova}, Yana and {Kulkarni}, Girish and {Loiacono}, Federica and {Lupi}, Alessandro and {Mazzucchelli}, Chiara and {Pan}, Zhiwei and {Rojas-Ruiz}, Sof{\'\i}a and {Strauss}, Michael A. and {Tee}, Wei Leong and {Trakhtenbrot}, Benny and {Trebitsch}, Maxime and {Venemans}, Bram and {Vestergaard}, Marianne and {Volonteri}, Marta and {Walter}, Fabian and {Xie}, Zhang-Liang and {Yue}, Minghao and {Zhang}, Haowen and {Zhang}, Huanian and {Zou}, Siwei},
        title = "{A SPectroscopic Survey of Biased Halos in the Reionization Era (ASPIRE): A First Look at the Rest-frame Optical Spectra of z > 6.5 Quasars Using JWST}",
      journal = {\apjl},
         year = 2023,
        month = jul,
       volume = {951},
       number = {1},
          eid = {L5},
        pages = {L5},
          doi = {10.3847/2041-8213/acc9c8},
archivePrefix = {arXiv},
       eprint = {2304.09888},
 primaryClass = {astro-ph.GA},
       adsurl = {https://ui.adsabs.harvard.edu/abs/2023ApJ...951L...5Y}
}

@ARTICLE{Riechers+20,
       author = {{Riechers}, Dominik A. and {Hodge}, Jacqueline A. and {Pavesi}, Riccardo and {Daddi}, Emanuele and {Decarli}, Roberto and {Ivison}, Rob J. and {Sharon}, Chelsea E. and {Smail}, Ian and {Walter}, Fabian and {Aravena}, Manuel and {Capak}, Peter L. and {Carilli}, Christopher L. and {Cox}, Pierre and {Cunha}, Elisabete da and {Dannerbauer}, Helmut and {Dickinson}, Mark and {Neri}, Roberto and {Wagg}, Jeff},
        title = "{COLDz: A High Space Density of Massive Dusty Starburst Galaxies {\ensuremath{\sim}}1 Billion Years after the Big Bang}",
      journal = {\apj},
         year = 2020,
        month = jun,
       volume = {895},
       number = {2},
          eid = {81},
        pages = {81},
          doi = {10.3847/1538-4357/ab8c48},
archivePrefix = {arXiv},
       eprint = {2004.10204},
 primaryClass = {astro-ph.GA},
       adsurl = {https://ui.adsabs.harvard.edu/abs/2020ApJ...895...81R}
}

@ARTICLE{Williams+19,
       author = {{Williams}, Christina C. and {Labbe}, Ivo and {Spilker}, Justin and {Stefanon}, Mauro and {Leja}, Joel and {Whitaker}, Katherine and {Bezanson}, Rachel and {Narayanan}, Desika and {Oesch}, Pascal and {Weiner}, Benjamin},
        title = "{Discovery of a Dark, Massive, ALMA-only Galaxy at z {\ensuremath{\sim}} 5-6 in a Tiny 3 mm Survey}",
      journal = {\apj},
         year = 2019,
        month = oct,
       volume = {884},
       number = {2},
          eid = {154},
        pages = {154},
          doi = {10.3847/1538-4357/ab44aa},
archivePrefix = {arXiv},
       eprint = {1905.11996},
 primaryClass = {astro-ph.GA},
       adsurl = {https://ui.adsabs.harvard.edu/abs/2019ApJ...884..154W}
}

@ARTICLE{Capak+11,
       author = {{Capak}, Peter L. and {Riechers}, Dominik and {Scoville}, Nick Z. and {Carilli}, Chris and {Cox}, Pierre and {Neri}, Roberto and {Robertson}, Brant and {Salvato}, Mara and {Schinnerer}, Eva and {Yan}, Lin and {Wilson}, Grant W. and {Yun}, Min and {Civano}, Francesca and {Elvis}, Martin and {Karim}, Alexander and {Mobasher}, Bahram and {Staguhn}, Johannes G.},
        title = "{A massive protocluster of galaxies at a redshift of z\raisebox{-0.5ex}\textasciitilde5.3}",
      journal = {\nat},
         year = 2011,
        month = feb,
       volume = {470},
       number = {7333},
        pages = {233-235},
          doi = {10.1038/nature09681},
archivePrefix = {arXiv},
       eprint = {1101.3586},
 primaryClass = {astro-ph.CO},
       adsurl = {https://ui.adsabs.harvard.edu/abs/2011Natur.470..233C}
}

@ARTICLE{Liu+19a,
       author = {{Liu}, Daizhong and {Lang}, P. and {Magnelli}, B. and {Schinnerer}, E. and {Leslie}, S. and {Fudamoto}, Y. and {Bondi}, M. and {Groves}, B. and {Jim{\'e}nez-Andrade}, E. and {Harrington}, K. and {Karim}, A. and {Oesch}, P.~A. and {Sargent}, M. and {Vardoulaki}, E. and {B{\v{a}}descu}, T. and {Moser}, L. and {Bertoldi}, F. and {Battisti}, A. and {da Cunha}, E. and {Zavala}, J. and {Vaccari}, M. and {Davidzon}, I. and {Riechers}, D. and {Aravena}, M.},
        title = "{Automated Mining of the ALMA Archive in the COSMOS Field (A$^{3}$COSMOS). I. Robust ALMA Continuum Photometry Catalogs and Stellar Mass and Star Formation Properties for {\ensuremath{\sim}}700 Galaxies at z = 0.5-6}",
      journal = {\apjs},
         year = 2019,
        month = oct,
       volume = {244},
       number = {2},
          eid = {40},
        pages = {40},
          doi = {10.3847/1538-4365/ab42da},
archivePrefix = {arXiv},
       eprint = {1910.12872},
 primaryClass = {astro-ph.GA},
       adsurl = {https://ui.adsabs.harvard.edu/abs/2019ApJS..244...40L}
}

@ARTICLE{Xiao+25,
       author = {{Xiao}, Mengyuan and {Williams}, Christina C. and {Oesch}, Pascal A. and {Elbaz}, David and {Dessauges-Zavadsky}, Miroslava and {Marques-Chaves}, Rui and {Bing}, Longji and {Ji}, Zhiyuan and {Weibel}, Andrea and {Bezanson}, Rachel and {Brammer}, Gabriel and {Casey}, Caitlin and {Cloonan}, Aidan P. and {Daddi}, Emanuele and {Dayal}, Pratika and {Faisst}, Andreas L. and {Franx}, Marijn and {Glazebrook}, Karl and {Hutter}, Anne and {Kartaltepe}, Jeyhan S. and {Labbe}, Ivo and {Lagache}, Guilaine and {Lim}, Seunghwan and {Magnelli}, Benjamin and {Martinez}, Felix and {Maseda}, Michael V. and {Nanayakkara}, Themiya and {Schaerer}, Daniel and {Whitaker}, Katherine E.},
        title = "{PANORAMIC: Discovery of an ultra-massive grand-design spiral galaxy at z {\ensuremath{\sim}} 5.2}",
      journal = {\aap},
         year = 2025,
        month = apr,
       volume = {696},
          eid = {A156},
        pages = {A156},
          doi = {10.1051/0004-6361/202453487},
archivePrefix = {arXiv},
       eprint = {2412.13264},
 primaryClass = {astro-ph.GA},
       adsurl = {https://ui.adsabs.harvard.edu/abs/2025A&A...696A.156X}
}

@ARTICLE{Lagache+25,
       author = {{Lagache}, G. and {Xiao}, M. and {Beelen}, A. and {Berta}, S. and {Ciesla}, L. and {Neri}, R. and {Pello}, R. and {Adam}, R. and {Ade}, P. and {Ajeddig}, H. and {Amarantidis}, S. and {Andr{\'e}}, P. and {Aussel}, H. and {Beno{\^\i}t}, A. and {B{\'e}thermin}, M. and {Bing}, L. -J. and {Bongiovanni}, A. and {Bounmy}, J. and {Bourrion}, O. and {Calvo}, M. and {Catalano}, A. and {Ch{\'e}rouvrier}, D. and {Chowdhury}, U. and {De Petris}, M. and {D{\'e}sert}, F. -X. and {Doyle}, S. and {Driessen}, E.~F.~C. and {Ejlali}, G. and {Ferragamo}, A. and {Gomez}, A. and {Goupy}, J. and {Hanser}, C. and {Katsioli}, S. and {K{\'e}ruzor{\'e}}, F. and {Kramer}, C. and {Ladjelate}, B. and {Leclercq}, S. and {Lestrade}, J. -F. and {Mac{\'\i}as-P{\'e}rez}, J.~F. and {Madden}, S.~C. and {Maury}, A. and {Mayet}, F. and {Monfardini}, A. and {Moyer-Anin}, A. and {Mu{\~n}oz-Echeverr{\'\i}a}, M. and {Myserlis}, I. and {Oesch}, P. and {Paliwal}, A. and {Perotto}, L. and {Pisano}, G. and {Ponthieu}, N. and {Rev{\'e}ret}, V. and {Rigby}, A.~J. and {Ritacco}, A. and {Roussel}, H. and {Ruppin}, F. and {S{\'a}nchez-Portal}, M. and {Savorgnano}, S. and {Schuster}, K. and {Sievers}, A. and {Tucker}, C. and {Zylka}, R.},
        title = "{Overdense fireworks in GOODS-N: Unveiling a record number of massive dusty star forming galaxies at z$\sim$5.2 with the N2CLS}",
      journal = {arXiv e-prints},
         year = 2025,
        month = jun,
          eid = {arXiv:2506.15322},
        pages = {arXiv:2506.15322},
          doi = {10.48550/arXiv.2506.15322},
archivePrefix = {arXiv},
       eprint = {2506.15322},
 primaryClass = {astro-ph.GA},
       adsurl = {https://ui.adsabs.harvard.edu/abs/2025arXiv250615322L}
}

@ARTICLE{Chabrier+03,
       author = {{Chabrier}, Gilles},
        title = "{Galactic Stellar and Substellar Initial Mass Function}",
      journal = {\pasp},
         year = 2003,
        month = jul,
       volume = {115},
       number = {809},
        pages = {763-795},
          doi = {10.1086/376392},
archivePrefix = {arXiv},
       eprint = {astro-ph/0304382},
 primaryClass = {astro-ph},
       adsurl = {https://ui.adsabs.harvard.edu/abs/2003PASP..115..763C}
}

@ARTICLE{[Planck+18,
       author = {{Planck Collaboration} and {Aghanim}, N. and {Akrami}, Y. and {Ashdown}, M. and {Aumont}, J. and {Baccigalupi}, C. and {Ballardini}, M. and {Banday}, A.~J. and {Barreiro}, R.~B. and {Bartolo}, N. and {Basak}, S. and {Battye}, R. and {Benabed}, K. and {Bernard}, J. -P. and {Bersanelli}, M. and {Bielewicz}, P. and {Bock}, J.~J. and {Bond}, J.~R. and {Borrill}, J. and {Bouchet}, F.~R. and {Boulanger}, F. and {Bucher}, M. and {Burigana}, C. and {Butler}, R.~C. and {Calabrese}, E. and {Cardoso}, J. -F. and {Carron}, J. and {Challinor}, A. and {Chiang}, H.~C. and {Chluba}, J. and {Colombo}, L.~P.~L. and {Combet}, C. and {Contreras}, D. and {Crill}, B.~P. and {Cuttaia}, F. and {de Bernardis}, P. and {de Zotti}, G. and {Delabrouille}, J. and {Delouis}, J. -M. and {Di Valentino}, E. and {Diego}, J.~M. and {Dor{\'e}}, O. and {Douspis}, M. and {Ducout}, A. and {Dupac}, X. and {Dusini}, S. and {Efstathiou}, G. and {Elsner}, F. and {En{\ss}lin}, T.~A. and {Eriksen}, H.~K. and {Fantaye}, Y. and {Farhang}, M. and {Fergusson}, J. and {Fernandez-Cobos}, R. and {Finelli}, F. and {Forastieri}, F. and {Frailis}, M. and {Fraisse}, A.~A. and {Franceschi}, E. and {Frolov}, A. and {Galeotta}, S. and {Galli}, S. and {Ganga}, K. and {G{\'e}nova-Santos}, R.~T. and {Gerbino}, M. and {Ghosh}, T. and {Gonz{\'a}lez-Nuevo}, J. and {G{\'o}rski}, K.~M. and {Gratton}, S. and {Gruppuso}, A. and {Gudmundsson}, J.~E. and {Hamann}, J. and {Handley}, W. and {Hansen}, F.~K. and {Herranz}, D. and {Hildebrandt}, S.~R. and {Hivon}, E. and {Huang}, Z. and {Jaffe}, A.~H. and {Jones}, W.~C. and {Karakci}, A. and {Keih{\"a}nen}, E. and {Keskitalo}, R. and {Kiiveri}, K. and {Kim}, J. and {Kisner}, T.~S. and {Knox}, L. and {Krachmalnicoff}, N. and {Kunz}, M. and {Kurki-Suonio}, H. and {Lagache}, G. and {Lamarre}, J. -M. and {Lasenby}, A. and {Lattanzi}, M. and {Lawrence}, C.~R. and {Le Jeune}, M. and {Lemos}, P. and {Lesgourgues}, J. and {Levrier}, F. and {Lewis}, A. and {Liguori}, M. and {Lilje}, P.~B. and {Lilley}, M. and {Lindholm}, V. and {L{\'o}pez-Caniego}, M. and {Lubin}, P.~M. and {Ma}, Y. -Z. and {Mac{\'\i}as-P{\'e}rez}, J.~F. and {Maggio}, G. and {Maino}, D. and {Mandolesi}, N. and {Mangilli}, A. and {Marcos-Caballero}, A. and {Maris}, M. and {Martin}, P.~G. and {Martinelli}, M. and {Mart{\'\i}nez-Gonz{\'a}lez}, E. and {Matarrese}, S. and {Mauri}, N. and {McEwen}, J.~D. and {Meinhold}, P.~R. and {Melchiorri}, A. and {Mennella}, A. and {Migliaccio}, M. and {Millea}, M. and {Mitra}, S. and {Miville-Desch{\^e}nes}, M. -A. and {Molinari}, D. and {Montier}, L. and {Morgante}, G. and {Moss}, A. and {Natoli}, P. and {N{\o}rgaard-Nielsen}, H.~U. and {Pagano}, L. and {Paoletti}, D. and {Partridge}, B. and {Patanchon}, G. and {Peiris}, H.~V. and {Perrotta}, F. and {Pettorino}, V. and {Piacentini}, F. and {Polastri}, L. and {Polenta}, G. and {Puget}, J. -L. and {Rachen}, J.~P. and {Reinecke}, M. and {Remazeilles}, M. and {Renzi}, A. and {Rocha}, G. and {Rosset}, C. and {Roudier}, G. and {Rubi{\~n}o-Mart{\'\i}n}, J.~A. and {Ruiz-Granados}, B. and {Salvati}, L. and {Sandri}, M. and {Savelainen}, M. and {Scott}, D. and {Shellard}, E.~P.~S. and {Sirignano}, C. and {Sirri}, G. and {Spencer}, L.~D. and {Sunyaev}, R. and {Suur-Uski}, A. -S. and {Tauber}, J.~A. and {Tavagnacco}, D. and {Tenti}, M. and {Toffolatti}, L. and {Tomasi}, M. and {Trombetti}, T. and {Valenziano}, L. and {Valiviita}, J. and {Van Tent}, B. and {Vibert}, L. and {Vielva}, P. and {Villa}, F. and {Vittorio}, N. and {Wandelt}, B.~D. and {Wehus}, I.~K. and {White}, M. and {White}, S.~D.~M. and {Zacchei}, A. and {Zonca}, A.},
        title = "{Planck 2018 results. VI. Cosmological parameters}",
      journal = {\aap},
         year = 2020,
        month = sep,
       volume = {641},
          eid = {A6},
        pages = {A6},
          doi = {10.1051/0004-6361/201833910},
archivePrefix = {arXiv},
       eprint = {1807.06209},
 primaryClass = {astro-ph.CO},
       adsurl = {https://ui.adsabs.harvard.edu/abs/2020A&A...641A...6P}
}

@ARTICLE{Fujimoto+25,
       author = {{Fujimoto}, Seiji and {Bezanson}, Rachel and {Labbe}, Ivo and {Brammer}, Gabriel and {Price}, Sedona H. and {Wang}, Bingjie and {Weaver}, John R. and {Fudamoto}, Yoshinobu and {Oesch}, Pascal A. and {Williams}, Christina C. and {Dayal}, Pratika and {Feldmann}, Robert and {Greene}, Jenny E. and {Leja}, Joel and {Whitaker}, Katherine E. and {Zitrin}, Adi and {Cutler}, Sam E. and {Furtak}, Lukas J. and {Pan}, Richard and {Chemerynska}, Iryna and {Kokorev}, Vasily and {Miller}, Tim B. and {Atek}, Hakim and {van Dokkum}, Pieter and {Juneau}, St{\'e}phanie and {Kassin}, Susan and {Khullar}, Gourav and {Marchesini}, Danilo and {Maseda}, Michael and {Nelson}, Erica J. and {Setton}, David J. and {Smit}, Renske},
        title = "{DUALZ{\textemdash}Deep UNCOVER-ALMA Legacy High-Z Survey}",
      journal = {\apjs},
         year = 2025,
        month = jun,
       volume = {278},
       number = {2},
          eid = {45},
        pages = {45},
          doi = {10.3847/1538-4365/adc677},
archivePrefix = {arXiv},
       eprint = {2309.07834},
 primaryClass = {astro-ph.GA},
       adsurl = {https://ui.adsabs.harvard.edu/abs/2025ApJS..278...45F}
}

@ARTICLE{Hodge+25,
       author = {{Hodge}, J.~A. and {da Cunha}, E. and {Kendrew}, S. and {Li}, J. and {Smail}, I. and {Westoby}, B.~A. and {Nayak}, O. and {Swinbank}, A.~M. and {Chen}, C. -C. and {Walter}, F. and {van der Werf}, P. and {Cracraft}, M. and {Battisti}, A. and {Brandt}, W.~N. and {Calistro Rivera}, G. and {Chapman}, S.~C. and {Cox}, P. and {Dannerbauer}, H. and {Decarli}, R. and {Frias Castillo}, M. and {Greve}, T.~R. and {Knudsen}, K.~K. and {Leslie}, S. and {Menten}, K.~M. and {Rybak}, M. and {Schinnerer}, E. and {Wardlow}, J.~L. and {Weiss}, A.},
        title = "{ALESS-JWST: Joint (Sub)kiloparsec JWST and ALMA Imaging of z \raisebox{-0.5ex}\textasciitilde 3 Submillimeter Galaxies Reveals Heavily Obscured Bulge Formation Events}",
      journal = {\apj},
         year = 2025,
        month = jan,
       volume = {978},
       number = {2},
          eid = {165},
        pages = {165},
          doi = {10.3847/1538-4357/ad9a52},
archivePrefix = {arXiv},
       eprint = {2407.15846},
 primaryClass = {astro-ph.GA},
       adsurl = {https://ui.adsabs.harvard.edu/abs/2025ApJ...978..165H}
}

@ARTICLE{Ikarashi+15,
       author = {{Ikarashi}, Soh and {Ivison}, R.~J. and {Caputi}, Karina I. and {Aretxaga}, Itziar and {Dunlop}, James S. and {Hatsukade}, Bunyo and {Hughes}, David H. and {Iono}, Daisuke and {Izumi}, Takuma and {Kawabe}, Ryohei and {Kohno}, Kotaro and {Lagos}, Claudia D.~P. and {Motohara}, Kentaro and {Nakanishi}, Kouichiro and {Ohta}, Kouji and {Tamura}, Yoichi and {Umehata}, Hideki and {Wilson}, Grant W. and {Yabe}, Kiyoto and {Yun}, Min S.},
        title = "{Compact Starbursts in z {\ensuremath{\sim}} 3-6 Submillimeter Galaxies Revealed by ALMA}",
      journal = {\apj},
         year = 2015,
        month = sep,
       volume = {810},
       number = {2},
          eid = {133},
        pages = {133},
          doi = {10.1088/0004-637X/810/2/133},
archivePrefix = {arXiv},
       eprint = {1411.5038},
 primaryClass = {astro-ph.GA},
       adsurl = {https://ui.adsabs.harvard.edu/abs/2015ApJ...810..133I}
}

@INPROCEEDINGS{Carpenter+23,
       author = {{Carpenter}, John and {Brogan}, Crystal and {Iono}, Daisuke and {Mroczkowski}, Tony},
        title = "{The ALMA Wideband Sensitivity Upgrade}",
    booktitle = {Physics and Chemistry of Star Formation: The Dynamical ISM Across Time and Spatial Scales},
         year = 2023,
       editor = {{Ossenkopf-Okada}, V. and {Schaaf}, R. and {Breloy}, I. and {Stutzki}, J.},
        month = feb,
        pages = {304},
          doi = {10.48550/arXiv.2211.00195},
archivePrefix = {arXiv},
       eprint = {2211.00195},
 primaryClass = {astro-ph.IM},
       adsurl = {https://ui.adsabs.harvard.edu/abs/2023pcsf.conf..304C}
}

@INPROCEEDINGS{Klaassen+19,
       author = {{Klaassen}, Pamela and {Mroczkowski}, Tony and {Bryan}, Sean and {Groppi}, Christopher and {Basu}, Kaustuv and {Cicone}, Claudia and {Dannerbauer}, Helmut and {De Breuck}, Carlos and {Fischer}, William J. and {Geach}, James and {Hatziminaoglou}, Evanthia and {Holland}, Wayne and {Kawabe}, Ryohei and {Sehgal}, Neelima and {Stanke}, Thomas and {van Kampen}, Eelco},
        title = "{The Atacama Large Aperture Submillimeter Telescope (AtLAST)}",
    booktitle = {Bulletin of the American Astronomical Society},
         year = 2019,
       volume = {51},
        month = sep,
          eid = {58},
        pages = {58},
          doi = {10.48550/arXiv.1907.04756},
archivePrefix = {arXiv},
       eprint = {1907.04756},
 primaryClass = {astro-ph.IM},
       adsurl = {https://ui.adsabs.harvard.edu/abs/2019BAAS...51g..58K}
}

@ARTICLE{Lovell+21,
       author = {{Lovell}, Christopher C. and {Vijayan}, Aswin P. and {Thomas}, Peter A. and {Wilkins}, Stephen M. and {Barnes}, David J. and {Irodotou}, Dimitrios and {Roper}, Will},
        title = "{First Light And Reionization Epoch Simulations (FLARES) - I. Environmental dependence of high-redshift galaxy evolution}",
      journal = {\mnras},
         year = 2021,
        month = jan,
       volume = {500},
       number = {2},
        pages = {2127-2145},
          doi = {10.1093/mnras/staa3360},
archivePrefix = {arXiv},
       eprint = {2004.07283},
 primaryClass = {astro-ph.GA},
       adsurl = {https://ui.adsabs.harvard.edu/abs/2021MNRAS.500.2127L}
}

@software{Arnouts+11,
       author = {{Arnouts}, S. and {Ilbert}, O.},
        title = "{LePHARE: Photometric Analysis for Redshift Estimate}",
 howpublished = {Astrophysics Source Code Library, record ascl:1108.009},
         year = 2011,
        month = aug,
          eid = {ascl:1108.009},
       adsurl = {https://ui.adsabs.harvard.edu/abs/2011ascl.soft08009A}
}

@ARTICLE{Yang+22,
       author = {{Yang}, Guang and {Boquien}, M{\'e}d{\'e}ric and {Brandt}, W.~N. and {Buat}, V{\'e}ronique and {Burgarella}, Denis and {Ciesla}, Laure and {Lehmer}, Bret D. and {Ma{\l}ek}, Katarzyna and {Mountrichas}, George and {Papovich}, Casey and {Pons}, Estelle and {Stalevski}, Marko and {Theul{\'e}}, Patrice and {Zhu}, Shifu},
        title = "{Fitting AGN/Galaxy X-Ray-to-radio SEDs with CIGALE and Improvement of the Code}",
      journal = {\apj},
         year = 2022,
        month = mar,
       volume = {927},
       number = {2},
          eid = {192},
        pages = {192},
          doi = {10.3847/1538-4357/ac4971},
archivePrefix = {arXiv},
       eprint = {2201.03718},
 primaryClass = {astro-ph.GA},
       adsurl = {https://ui.adsabs.harvard.edu/abs/2022ApJ...927..192Y}
}

@ARTICLE{Boquien+19,
       author = {{Boquien}, M. and {Burgarella}, D. and {Roehlly}, Y. and {Buat}, V. and {Ciesla}, L. and {Corre}, D. and {Inoue}, A.~K. and {Salas}, H.},
        title = "{CIGALE: a python Code Investigating GALaxy Emission}",
      journal = {\aap},
         year = 2019,
        month = feb,
       volume = {622},
          eid = {A103},
        pages = {A103},
          doi = {10.1051/0004-6361/201834156},
archivePrefix = {arXiv},
       eprint = {1811.03094},
 primaryClass = {astro-ph.GA},
       adsurl = {https://ui.adsabs.harvard.edu/abs/2019A&A...622A.103B}
}

@ARTICLE{Baker+25,
       author = {{Baker}, William M. and {Lim}, Seunghwan and {D'Eugenio}, Francesco and {Maiolino}, Roberto and {Ji}, Zhiyuan and {Arribas}, Santiago and {Bunker}, Andrew J. and {Carniani}, Stefano and {Charlot}, Stephane and {de Graaff}, Anna and {Hainline}, Kevin and {Looser}, Tobias J. and {Lyu}, Jianwei and {Rinaldi}, Pierluigi and {Robertson}, Brant and {Schaller}, Matthieu and {Schaye}, Joop and {Scholtz}, Jan and {{\"U}bler}, Hannah and {Williams}, Christina C. and {Willmer}, Christopher N.~A. and {Willott}, Chris and {Zhu}, Yongda},
        title = "{The abundance and nature of high-redshift quiescent galaxies from JADES spectroscopy and the FLAMINGO simulations}",
      journal = {\mnras},
         year = 2025,
        month = may,
       volume = {539},
       number = {1},
        pages = {557-589},
          doi = {10.1093/mnras/staf475},
archivePrefix = {arXiv},
       eprint = {2410.14773},
 primaryClass = {astro-ph.GA},
       adsurl = {https://ui.adsabs.harvard.edu/abs/2025MNRAS.539..557B}
}

@ARTICLE{Song+20,
       author = {{Song}, Hyunmi and {Seon}, Kwang-Il and {Hwang}, Ho Seong},
        title = "{Ly{\ensuremath{\alpha}} Radiative Transfer: Modeling Spectrum and Surface Brightness Profiles of Ly{\ensuremath{\alpha}}-emitting Galaxies at Z = 3-6}",
      journal = {\apj},
         year = 2020,
        month = sep,
       volume = {901},
       number = {1},
          eid = {41},
        pages = {41},
          doi = {10.3847/1538-4357/abac02},
archivePrefix = {arXiv},
       eprint = {2007.08172},
 primaryClass = {astro-ph.GA},
       adsurl = {https://ui.adsabs.harvard.edu/abs/2020ApJ...901...41S}
}

@ARTICLE{Liao+19,
       author = {{Liao}, Shihong and {Gao}, Liang},
        title = "{Impact of filaments on galaxy formation in their residing dark matter haloes}",
      journal = {\mnras},
         year = 2019,
        month = may,
       volume = {485},
       number = {1},
        pages = {464-473},
          doi = {10.1093/mnras/stz441},
archivePrefix = {arXiv},
       eprint = {1805.10944},
 primaryClass = {astro-ph.GA},
       adsurl = {https://ui.adsabs.harvard.edu/abs/2019MNRAS.485..464L}
}

@ARTICLE{DiMatteo+17,
       author = {{Di Matteo}, Tiziana and {Croft}, Rupert A.~C. and {Feng}, Yu and {Waters}, Dacen and {Wilkins}, Stephen},
        title = "{The origin of the most massive black holes at high-z: BlueTides and the next quasar frontier}",
      journal = {\mnras},
         year = 2017,
        month = jun,
       volume = {467},
       number = {4},
        pages = {4243-4251},
          doi = {10.1093/mnras/stx319},
archivePrefix = {arXiv},
       eprint = {1606.08871},
 primaryClass = {astro-ph.GA},
       adsurl = {https://ui.adsabs.harvard.edu/abs/2017MNRAS.467.4243D}
}

@ARTICLE{Dekel+13,
       author = {{Dekel}, A. and {Zolotov}, A. and {Tweed}, D. and {Cacciato}, M. and {Ceverino}, D. and {Primack}, J.~R.},
        title = "{Toy models for galaxy formation versus simulations}",
      journal = {\mnras},
         year = 2013,
        month = oct,
       volume = {435},
       number = {2},
        pages = {999-1019},
          doi = {10.1093/mnras/stt1338},
archivePrefix = {arXiv},
       eprint = {1303.3009},
 primaryClass = {astro-ph.CO},
       adsurl = {https://ui.adsabs.harvard.edu/abs/2013MNRAS.435..999D}
}

@ARTICLE{Bethermin+13,
       author = {{B{\'e}thermin}, Matthieu and {Wang}, Lingyu and {Dor{\'e}}, Olivier and {Lagache}, Guilaine and {Sargent}, Mark and {Daddi}, Emanuele and {Cousin}, Morgane and {Aussel}, Herv{\'e}},
        title = "{The redshift evolution of the distribution of star formation among dark matter halos as seen in the infrared}",
      journal = {\aap},
         year = 2013,
        month = sep,
       volume = {557},
          eid = {A66},
        pages = {A66},
          doi = {10.1051/0004-6361/201321688},
archivePrefix = {arXiv},
       eprint = {1304.3936},
 primaryClass = {astro-ph.CO},
       adsurl = {https://ui.adsabs.harvard.edu/abs/2013A&A...557A..66B}
}

@ARTICLE{Chiang+13,
       author = {{Chiang}, Yi-Kuan and {Overzier}, Roderik and {Gebhardt}, Karl},
        title = "{Ancient Light from Young Cosmic Cities: Physical and Observational Signatures of Galaxy Proto-clusters}",
      journal = {\apj},
         year = 2013,
        month = dec,
       volume = {779},
       number = {2},
          eid = {127},
        pages = {127},
          doi = {10.1088/0004-637X/779/2/127},
archivePrefix = {arXiv},
       eprint = {1310.2938},
 primaryClass = {astro-ph.CO},
       adsurl = {https://ui.adsabs.harvard.edu/abs/2013ApJ...779..127C}
}

@ARTICLE{Calvi+21,
       author = {{Calvi}, Rosa and {Dannerbauer}, Helmut and {Arrabal Haro}, Pablo and {Rodr{\'\i}guez Espinosa}, Jos{\'e} M. and {Mu{\~n}oz-Tu{\~n}{\'o}n}, Casiana and {P{\'e}rez Gonz{\'a}lez}, Pablo G. and {Geier}, Stefan},
        title = "{Probing the existence of a rich galaxy overdensity at z = 5.2}",
      journal = {\mnras},
         year = 2021,
        month = apr,
       volume = {502},
       number = {3},
        pages = {4558-4575},
          doi = {10.1093/mnras/staa4037},
archivePrefix = {arXiv},
       eprint = {2101.02747},
 primaryClass = {astro-ph.GA},
       adsurl = {https://ui.adsabs.harvard.edu/abs/2021MNRAS.502.4558C}
}

@ARTICLE{Brinch+23,
       author = {{Brinch}, Malte and {Greve}, Thomas R. and {Weaver}, John R. and {Brammer}, Gabriel and {Ilbert}, Olivier and {Shuntov}, Marko and {Jin}, Shuowen and {Liu}, Daizhong and {Gim{\'e}nez-Arteaga}, Clara and {Casey}, Caitlin M. and {Davidson}, Iary and {Fujimoto}, Seiji and {Koekemoer}, Anton M. and {Kokorev}, Vasily and {Magdis}, Georgios and {McCracken}, H.~J. and {McPartland}, Conor J.~R. and {Mobasher}, Bahram and {Sanders}, David B. and {Toft}, Sune and {Valentino}, Francesco and {Zamorani}, Giovanni and {Zavala}, Jorge and {Cosmos Team}},
        title = "{COSMOS2020: Identification of High-z Protocluster Candidates in COSMOS}",
      journal = {\apj},
         year = 2023,
        month = feb,
       volume = {943},
       number = {2},
          eid = {153},
        pages = {153},
          doi = {10.3847/1538-4357/ac9d96},
archivePrefix = {arXiv},
       eprint = {2210.17334},
 primaryClass = {astro-ph.GA},
       adsurl = {https://ui.adsabs.harvard.edu/abs/2023ApJ...943..153B}
}

@ARTICLE{Wang+23,
       author = {{Wang}, Feige and {Yang}, Jinyi and {Hennawi}, Joseph F. and {Fan}, Xiaohui and {Sun}, Fengwu and {Champagne}, Jaclyn B. and {Costa}, Tiago and {Habouzit}, Melanie and {Endsley}, Ryan and {Li}, Zihao and {Lin}, Xiaojing and {Meyer}, Romain A. and {Schindler}, Jan-Torge and {Wu}, Yunjing and {Ba{\~n}ados}, Eduardo and {Barth}, Aaron J. and {Bhowmick}, Aklant K. and {Bieri}, Rebekka and {Blecha}, Laura and {Bosman}, Sarah and {Cai}, Zheng and {Colina}, Luis and {Connor}, Thomas and {Davies}, Frederick B. and {Decarli}, Roberto and {De Rosa}, Gisella and {Drake}, Alyssa B. and {Egami}, Eiichi and {Eilers}, Anna-Christina and {Evans}, Analis E. and {Farina}, Emanuele Paolo and {Haiman}, Zoltan and {Jiang}, Linhua and {Jin}, Xiangyu and {Jun}, Hyunsung D. and {Kakiichi}, Koki and {Khusanova}, Yana and {Kulkarni}, Girish and {Li}, Mingyu and {Liu}, Weizhe and {Loiacono}, Federica and {Lupi}, Alessandro and {Mazzucchelli}, Chiara and {Onoue}, Masafusa and {Pudoka}, Maria A. and {Rojas-Ruiz}, Sof{\'\i}a and {Shen}, Yue and {Strauss}, Michael A. and {Tee}, Wei Leong and {Trakhtenbrot}, Benny and {Trebitsch}, Maxime and {Venemans}, Bram and {Volonteri}, Marta and {Walter}, Fabian and {Xie}, Zhang-Liang and {Yue}, Minghao and {Zhang}, Haowen and {Zhang}, Huanian and {Zou}, Siwei},
        title = "{A SPectroscopic Survey of Biased Halos in the Reionization Era (ASPIRE): JWST Reveals a Filamentary Structure around a z = 6.61 Quasar}",
      journal = {\apjl},
         year = 2023,
        month = jul,
       volume = {951},
       number = {1},
          eid = {L4},
        pages = {L4},
          doi = {10.3847/2041-8213/accd6f},
archivePrefix = {arXiv},
       eprint = {2304.09894},
 primaryClass = {astro-ph.GA},
       adsurl = {https://ui.adsabs.harvard.edu/abs/2023ApJ...951L...4W}
}

@ARTICLE{Barrufet+23b,
       author = {{Barrufet}, L. and {Oesch}, P.~A. and {Bouwens}, R. and {Inami}, H. and {Sommovigo}, L. and {Algera}, H. and {da Cunha}, E. and {Aravena}, M. and {Dayal}, P. and {Ferrara}, A. and {Fudamoto}, Y. and {Gonzalez}, V. and {Graziani}, L. and {Hygate}, A.~P.~S. and {de Looze}, I. and {Nanayakkara}, T. and {Pallottini}, A. and {Schneider}, R. and {Stefanon}, M. and {Topping}, M. and {van der Werf}, P.},
        title = "{The ALMA REBELS Survey: the first infrared luminosity function measurement at z {\ensuremath{\sim}} 7}",
      journal = {\mnras},
         year = 2023,
        month = jul,
       volume = {522},
       number = {3},
        pages = {3926-3934},
          doi = {10.1093/mnras/stad1259},
archivePrefix = {arXiv},
       eprint = {2303.11321},
 primaryClass = {astro-ph.GA},
       adsurl = {https://ui.adsabs.harvard.edu/abs/2023MNRAS.522.3926B}
}

@ARTICLE{Zhang+25,
       author = {{Zhang}, Yunchong and {de Graaff}, Anna and {Setton}, David J. and {Price}, Sedona H. and {Bezanson}, Rachel and {Lagos}, Claudia del P. and {Cutler}, Sam E. and {McConachie}, Ian and {Cleri}, Nikko J. and {Cooper}, Olivia R. and {Gottumukkala}, Rashmi and {Greene}, Jenny E. and {Hirschmann}, Michaela and {Khullar}, Gourav and {Labbe}, Ivo and {Leja}, Joel and {Maseda}, Michael V. and {Matthee}, Jorryt and {Miller}, Tim B. and {Nanayakkara}, Themiya and {Suess}, Katherine A. and {Wang}, Bingjie and {Whitaker}, Katherine E. and {Williams}, Christina C.},
        title = "{RUBIES spectroscopically confirms the high number density of quiescent galaxies from $\mathbf{2<z<5}$}",
      journal = {arXiv e-prints},
         year = 2025,
        month = aug,
          eid = {arXiv:2508.08577},
        pages = {arXiv:2508.08577},
          doi = {10.48550/arXiv.2508.08577},
archivePrefix = {arXiv},
       eprint = {2508.08577},
 primaryClass = {astro-ph.GA},
       adsurl = {https://ui.adsabs.harvard.edu/abs/2025arXiv250808577Z}
}

@ARTICLE{Akins+25,
       author = {{Akins}, Hollis B. and {Casey}, Caitlin M. and {Champagne}, Jaclyn B. and {Cooper}, Olivia and {Franco}, Maximilien and {Fujimoto}, Seiji and {Knudsen}, Kirsten K. and {Koekemoer}, Anton M. and {Long}, Arianna S. and {Man}, Allison and {Manning}, Sinclaire M. and {McKinney}, Jed and {Zavala}, Jorge and {Arrabal Haro}, Pablo and {Dickinson}, Mark and {Kokorev}, Vasily and {Taylor}, Anthony J.},
        title = "{JWST+ALMA reveal the ISM kinematics and stellar structure of MAMBO-9, a merging pair of DSFGs in an overdense environment at $z=5.85$}",
      journal = {arXiv e-prints},
         year = 2025,
        month = aug,
          eid = {arXiv:2508.06607},
        pages = {arXiv:2508.06607},
          doi = {10.48550/arXiv.2508.06607},
archivePrefix = {arXiv},
       eprint = {2508.06607},
 primaryClass = {astro-ph.GA},
       adsurl = {https://ui.adsabs.harvard.edu/abs/2025arXiv250806607A}
}

@ARTICLE{Bethermin+25,
       author = {{B{\'e}thermin}, M. and {Lagache}, G. and {Carvajal-Bohorquez}, C. and {Adam}, R. and {Ade}, P. and {Ajeddig}, H. and {Amarantidis}, S. and {Andr{\'e}}, P. and {Aussel}, H. and {Beelen}, A. and {Beno{\^\i}t}, A. and {Berta}, S. and {Bing}, L.~J. and {Bongiovanni}, A. and {Bounmy}, J. and {Bourrion}, O. and {Calvo}, M. and {Catalano}, A. and {Ch{\'e}rouvrier}, D. and {De Petris}, M. and {D{\'e}sert}, F. -X. and {Doyle}, S. and {Driessen}, E.~F.~C. and {Ejlali}, G. and {Ferragamo}, A. and {Gomez}, A. and {Goupy}, J. and {Hanser}, C. and {Katsioli}, S. and {K{\'e}ruzor{\'e}}, F. and {Kramer}, C. and {Ladjelate}, B. and {Leclercq}, S. and {Lestrade}, J. -F. and {Mac{\'\i}as-P{\'e}rez}, J.~F. and {Madden}, S.~C. and {Maury}, A. and {Mayet}, F. and {Monfardini}, A. and {Moyer-Anin}, A. and {Mu{\~n}oz-Echeverr{\'\i}a}, M. and {Myserlis}, I. and {Paliwal}, A. and {Perotto}, L. and {Pisano}, G. and {Ponthieu}, N. and {Rev{\'e}ret}, V. and {Rigby}, A.~J. and {Ritacco}, A. and {Roussel}, H. and {Ruppin}, F. and {S{\'a}nchez-Portal}, M. and {Savorgnano}, S. and {Schuster}, K. and {Sievers}, A. and {Tucker}, C. and {Zylka}, R.},
        title = "{The NIKA2 cosmological legacy survey at 2 mm: catalogs, colors, redshift distributions, and implications for deep surveys}",
      journal = {arXiv e-prints},
         year = 2025,
        month = jun,
          eid = {arXiv:2506.22046},
        pages = {arXiv:2506.22046},
          doi = {10.48550/arXiv.2506.22046},
archivePrefix = {arXiv},
       eprint = {2506.22046},
 primaryClass = {astro-ph.GA},
       adsurl = {https://ui.adsabs.harvard.edu/abs/2025arXiv250622046B}
}

@ARTICLE{Harish+25,
       author = {{Harish}, Santosh and {Kartaltepe}, Jeyhan S. and {Liu}, Daizhong and {Koekemoer}, Anton M. and {Casey}, Caitlin M. and {Franco}, Maximilien and {Akins}, Hollis B. and {Ilbert}, Olivier and {Shuntov}, Marko and {Drakos}, Nicole E. and {Engesser}, Mike and {Faisst}, Andreas L. and {Gozaliasl}, Ghassem and {Martin}, Crystal L. and {Hirschmann}, Michaela and {Kokorev}, Vasily and {Lambrides}, Erini and {McCracken}, Henry Joy and {McKinney}, Jed and {Paquereau}, Louise and {Rhodes}, Jason and {Robertson}, Brant E.},
        title = "{COSMOS-Web: MIRI Data Reduction and Number Counts at 7.7$μ$m using JWST}",
      journal = {arXiv e-prints},
         year = 2025,
        month = jun,
          eid = {arXiv:2506.03306},
        pages = {arXiv:2506.03306},
          doi = {10.48550/arXiv.2506.03306},
archivePrefix = {arXiv},
       eprint = {2506.03306},
 primaryClass = {astro-ph.GA},
       adsurl = {https://ui.adsabs.harvard.edu/abs/2025arXiv250603306H}
}

@ARTICLE{Franco+25,
       author = {{Franco}, Maximilien and {Casey}, Caitlin M. and {Koekemoer}, Anton M. and {Liu}, Daizhong and {Bagley}, Micaela B. and {McCracken}, Henry Joy and {Kartaltepe}, Jeyhan S. and {Akins}, Hollis B. and {Ilbert}, Olivier and {Shuntov}, Marko and {Harish}, Santosh and {Robertson}, Brant E. and {Arango-Toro}, Rafael C. and {Battisti}, Andrew J. and {Chartab}, Nima and {Drakos}, Nicole E. and {Faisst}, Andreas L. and {Flayhart}, Carter and {Gozaliasl}, Ghassem and {Hirschmann}, Michaela and {Massey}, Richard and {Rhodes}, Jason and {Sattari}, Zahra and {Scognamiglio}, Diana and {Weaver}, John R. and {Yang}, Lilan and {Zavala}, Jorge A. and {Berman}, Edward M. and {Gentile}, Fabrizio and {Gillman}, Steven and {Long}, Arianna S. and {Magdis}, Georgios and {McCleary}, Jacqueline E. and {McKinney}, Jed and {Mobasher}, Bahram and {Paquereau}, Louise and {Rest}, Armin and {Sanders}, David B. and {Toft}, Sune and {Yu}, Si-Yue},
        title = "{COSMOS-Web: Comprehensive Data Reduction for Wide-Area JWST NIRCam Imaging}",
      journal = {arXiv e-prints},
         year = 2025,
        month = jun,
          eid = {arXiv:2506.03256},
        pages = {arXiv:2506.03256},
          doi = {10.48550/arXiv.2506.03256},
archivePrefix = {arXiv},
       eprint = {2506.03256},
 primaryClass = {astro-ph.IM},
       adsurl = {https://ui.adsabs.harvard.edu/abs/2025arXiv250603256F}
}

@ARTICLE{Shuntov+25,
       author = {{Shuntov}, Marko and {Akins}, Hollis B. and {Paquereau}, Louise and {Casey}, Caitlin M. and {Ilbert}, Olivier and {Arango-Toro}, Rafael C. and {McCracken}, Henry Joy and {Franco}, Maximilien and {Harish}, Santosh and {Kartaltepe}, Jeyhan S. and {Koekemoer}, Anton M. and {Yang}, Lilan and {Huertas-Company}, Marc and {Berman}, Edward M. and {McCleary}, Jacqueline E. and {Toft}, Sune and {Gavazzi}, Rapha{\"e}l and {Achenbach}, Mark J. and {Bertin}, Emmanuel and {Brinch}, Malte and {Champagne}, Jackie and {Chartab}, Nima and {Drakos}, Nicole E. and {Egami}, Eiichi and {Endsley}, Ryan and {Faisst}, Andreas L. and {Fan}, Xiaohui and {Flayhart}, Carter and {Hartley}, William G. and {Hatamnia}, Hossein and {Gozaliasl}, Ghassem and {Gentile}, Fabrizio and {Jermann}, Iris and {Jin}, Shuowen and {Kakiichi}, Koki and {Khostovan}, Ali Ahmad and {K{\"u}mmel}, Martin and {Laigle}, Clotilde and {Laishram}, Ronaldo and {Lambrides}, Erini and {Liu}, Daizhong and {Lyu}, Jianwei and {Magdis}, Georgios and {Mobasher}, Bahram and {Moutard}, Thibaud and {Renzini}, Alvio and {Robertson}, Brant E. and {Schefer}, Marc and {Scognamiglio}, Diana and {Scoville}, Nick and {Sattari}, Zahra and {Sanders}, David B. and {Taamoli}, Sina and {Trakhtenbrot}, Benny and {Valentino}, Francesco and {Wang}, Feige and {Weaver}, John R. and {Yang}, Jinyl},
        title = "{COSMOS2025: The COSMOS-Web galaxy catalog of photometry, morphology, redshifts, and physical parameters from JWST, HST, and ground-based imaging}",
      journal = {arXiv e-prints},
         year = 2025,
        month = jun,
          eid = {arXiv:2506.03243},
        pages = {arXiv:2506.03243},
          doi = {10.48550/arXiv.2506.03243},
archivePrefix = {arXiv},
       eprint = {2506.03243},
 primaryClass = {astro-ph.GA},
       adsurl = {https://ui.adsabs.harvard.edu/abs/2025arXiv250603243S}
}

@ARTICLE{Sun+25,
       author = {{Sun}, Fengwu and {Yang}, Jinyi and {Wang}, Feige and {Eisenstein}, Daniel J. and {Decarli}, Roberto and {Fan}, Xiaohui and {Rieke}, George H. and {Ba{\~n}ados}, Eduardo and {Bosman}, Sarah E.~I. and {Cai}, Zheng and {Champagne}, Jaclyn B. and {Colina}, Luis and {D'Eugenio}, Francesco and {Fudamoto}, Yoshinobu and {Li}, Mingyu and {Lin}, Xiaojing and {Liu}, Weizhe and {Lyu}, Jianwei and {Mazzucchelli}, Chiara and {Jin}, Xiangyu and {Jun}, Hyunsung D. and {Wu}, Yunjing and {Zhang}, Huanian},
        title = "{The Identification of Two JWST/NIRCam-Dark Starburst Galaxies at $z=6.6$ with ALMA}",
      journal = {arXiv e-prints},
         year = 2025,
        month = jun,
          eid = {arXiv:2506.06418},
        pages = {arXiv:2506.06418},
          doi = {10.48550/arXiv.2506.06418},
archivePrefix = {arXiv},
       eprint = {2506.06418},
 primaryClass = {astro-ph.GA},
       adsurl = {https://ui.adsabs.harvard.edu/abs/2025arXiv250606418S}
}

@ARTICLE{Berta+25,
       author = {{Berta}, S. and {Lagache}, G. and {Beelen}, A. and {Adam}, R. and {Ade}, P. and {Ajeddig}, H. and {Amarantidis}, S. and {Andr{\'e}}, P. and {Aussel}, H. and {Beno{\^\i}t}, A. and {Bethermin}, M. and {Bing}, L. -J. and {Bongiovanni}, A. and {Bounmy}, J. and {Bourrion}, O. and {Calvo}, M. and {Catalano}, A. and {Ch{\'e}rouvrier}, D. and {Ciesla}, L. and {De Petris}, M. and {D{\'e}sert}, F. -X. and {Doyle}, S. and {Driessen}, E.~F.~C. and {Ejlali}, G. and {Elbaz}, D. and {Ferragamo}, A. and {Gomez}, A. and {Goupy}, J. and {Hanser}, C. and {Katsioli}, S. and {K{\'e}ruzor{\'e}}, F. and {Kramer}, C. and {Ladjelate}, B. and {Leclercq}, S. and {Lestrade}, J. -F. and {Mac{\'\i}as-P{\'e}rez}, J.~F. and {Madden}, S.~C. and {Maury}, A. and {Mayet}, F. and {Messias}, H. and {Monfardini}, A. and {Moyer-Anin}, A. and {Mu{\~n}oz-Echeverr{\'\i}a}, M. and {Myserlis}, I. and {Neri}, R. and {Paliwal}, A. and {Perotto}, L. and {Pisano}, G. and {Ponthieu}, N. and {Rev{\'e}ret}, V. and {Rigby}, A.~J. and {Ritacco}, A. and {Roussel}, H. and {Ruppin}, F. and {S{\'a}nchez-Portal}, M. and {Savorgnano}, S. and {Schuster}, K. and {Sievers}, A. and {Tucker}, C. and {Xiao}, M. -Y. and {Zylka}, R.},
        title = "{A panchromatic view of N2CLS GOODS-N: The evolution of the dust cosmic density since z {\ensuremath{\sim}} 7}",
      journal = {\aap},
         year = 2025,
        month = apr,
       volume = {696},
          eid = {A193},
        pages = {A193},
          doi = {10.1051/0004-6361/202452894},
archivePrefix = {arXiv},
       eprint = {2503.07706},
 primaryClass = {astro-ph.GA},
       adsurl = {https://ui.adsabs.harvard.edu/abs/2025A&A...696A.193B}
}

@ARTICLE{Weibel+25,
       author = {{Weibel}, Andrea and {de Graaff}, Anna and {Setton}, David J. and {Miller}, Tim B. and {Oesch}, Pascal A. and {Brammer}, Gabriel and {Lagos}, Claudia D.~P. and {Whitaker}, Katherine E. and {Williams}, Christina C. and {Baggen}, Josephine F.~W. and {Bezanson}, Rachel and {Boogaard}, Leindert A. and {Cleri}, Nikko J. and {Greene}, Jenny E. and {Hirschmann}, Michaela and {Hviding}, Raphael E. and {Kuruvanthodi}, Adarsh and {Labb{\'e}}, Ivo and {Leja}, Joel and {Maseda}, Michael V. and {Matthee}, Jorryt and {McConachie}, Ian and {Naidu}, Rohan P. and {Roberts-Borsani}, Guido and {Schaerer}, Daniel and {Suess}, Katherine A. and {Valentino}, Francesco and {van Dokkum}, Pieter and {Wang}, Bingjie},
        title = "{RUBIES Reveals a Massive Quiescent Galaxy at z = 7.3}",
      journal = {\apj},
         year = 2025,
        month = apr,
       volume = {983},
       number = {1},
          eid = {11},
        pages = {11},
          doi = {10.3847/1538-4357/adab7a},
archivePrefix = {arXiv},
       eprint = {2409.03829},
 primaryClass = {astro-ph.GA},
       adsurl = {https://ui.adsabs.harvard.edu/abs/2025ApJ...983...11W}
}

@ARTICLE{Champagne+25,
       author = {{Champagne}, Jaclyn B. and {Wang}, Feige and {Zhang}, Haowen and {Yang}, Jinyi and {Fan}, Xiaohui and {Hennawi}, Joseph F. and {Sun}, Fengwu and {Ba{\~n}ados}, Eduardo and {Bosman}, Sarah E.~I. and {Costa}, Tiago and {Eilers}, Anna-Christina and {Endsley}, Ryan and {Jin}, Xiangyu and {Jun}, Hyunsung D. and {Li}, Mingyu and {Lin}, Xiaojing and {Liu}, Weizhe and {Loiacono}, Federica and {Lupi}, Alessandro and {Mazzucchelli}, Chiara and {Pudoka}, Maria and {Protu{\v{s}}ov{\`a}}, Klaudia and {Rojas-Ruiz}, Sof{\'\i}a and {Tee}, Wei Leong and {Trebitsch}, Maxime and {Venemans}, Bram P. and {Zhuang}, Ming-Yang and {Zou}, Siwei},
        title = "{A Quasar-anchored Protocluster at z = 6.6 in the ASPIRE Survey. I. Properties of [O III] Emitters in a 10 Mpc Overdensity Structure}",
      journal = {\apj},
         year = 2025,
        month = mar,
       volume = {981},
       number = {2},
          eid = {113},
        pages = {113},
          doi = {10.3847/1538-4357/adb1bd},
archivePrefix = {arXiv},
       eprint = {2410.03826},
 primaryClass = {astro-ph.GA},
       adsurl = {https://ui.adsabs.harvard.edu/abs/2025ApJ...981..113C}
}

@ARTICLE{Champagne+25b,
       author = {{Champagne}, Jaclyn B. and {Wang}, Feige and {Yang}, Jinyi and {Fan}, Xiaohui and {Hennawi}, Joseph F. and {Sun}, Fengwu and {Ba{\~n}ados}, Eduardo and {Bosman}, Sarah E.~I. and {Costa}, Tiago and {Habouzit}, Melanie and {Jin}, Xiangyu and {Jun}, Hyunsung D. and {Li}, Mingyu and {Liu}, Weizhe and {Loiacono}, Federica and {Lupi}, Alessandro and {Mazzucchelli}, Chiara and {Pudoka}, Maria and {Rojas-Ruiz}, Sof{\'\i}a and {Tee}, Wei Leong and {Trebitsch}, Maxime and {Zhang}, Haowen and {Zhuang}, Ming-Yang and {Zou}, Siwei},
        title = "{A Quasar-anchored Protocluster at z = 6.6 in the ASPIRE Survey. II. An Environmental Analysis of Galaxy Properties in an Overdense Structure}",
      journal = {\apj},
         year = 2025,
        month = mar,
       volume = {981},
       number = {2},
          eid = {114},
        pages = {114},
          doi = {10.3847/1538-4357/adb1bc},
archivePrefix = {arXiv},
       eprint = {2410.03827},
 primaryClass = {astro-ph.GA},
       adsurl = {https://ui.adsabs.harvard.edu/abs/2025ApJ...981..114C}
}

@ARTICLE{deGraaff+25,
       author = {{de Graaff}, Anna and {Setton}, David J. and {Brammer}, Gabriel and {Cutler}, Sam and {Suess}, Katherine A. and {Labb{\'e}}, Ivo and {Leja}, Joel and {Weibel}, Andrea and {Maseda}, Michael V. and {Whitaker}, Katherine E. and {Bezanson}, Rachel and {Boogaard}, Leindert A. and {Cleri}, Nikko J. and {De Lucia}, Gabriella and {Franx}, Marijn and {Greene}, Jenny E. and {Hirschmann}, Michaela and {Matthee}, Jorryt and {McConachie}, Ian and {Naidu}, Rohan P. and {Oesch}, Pascal A. and {Price}, Sedona H. and {Rix}, Hans-Walter and {Valentino}, Francesco and {Wang}, Bingjie and {Williams}, Christina C.},
        title = "{Efficient formation of a massive quiescent galaxy at redshift 4.9}",
      journal = {Nature Astronomy},
         year = 2025,
        month = feb,
       volume = {9},
        pages = {280-292},
          doi = {10.1038/s41550-024-02424-3},
archivePrefix = {arXiv},
       eprint = {2404.05683},
 primaryClass = {astro-ph.GA},
       adsurl = {https://ui.adsabs.harvard.edu/abs/2025NatAs...9..280D}
}

@ARTICLE{Herard-Demanche+25,
       author = {{Herard-Demanche}, Thomas and {Bouwens}, Rychard J. and {Oesch}, Pascal A. and {Naidu}, Rohan P. and {Decarli}, Roberto and {Nelson}, Erica J. and {Brammer}, Gabriel and {Weibel}, Andrea and {Xiao}, Mengyuan and {Stefanon}, Mauro and {Walter}, Fabian and {Matthee}, Jorryt and {Meyer}, Romain A. and {Wuyts}, Stijn and {Reddy}, Naveen and {Rowland}, Lucie and {van Leeuwen}, Ivana and {Haro}, Pablo Arrabal and {Dannerbauer}, Helmut and {Shapley}, Alice E. and {Chisholm}, John and {van Dokkum}, Pieter and {Labbe}, Ivo and {Illingworth}, Garth and {Schaerer}, Daniel and {Shivaei}, Irene},
        title = "{Mapping dusty galaxy growth at z > 5 with FRESCO: detection of H{\ensuremath{\alpha}} in submm galaxy HDF850.1 and the surrounding overdense structures}",
      journal = {\mnras},
         year = 2025,
        month = feb,
       volume = {537},
       number = {2},
        pages = {788-808},
          doi = {10.1093/mnras/staf030},
archivePrefix = {arXiv},
       eprint = {2309.04525},
 primaryClass = {astro-ph.GA},
       adsurl = {https://ui.adsabs.harvard.edu/abs/2025MNRAS.537..788H}
}

@ARTICLE{delaVega+25,
       author = {{de la Vega}, Alexander and {Babcock}, Mitchell D. and {Mobasher}, Bahram and {Riemann}, Dominik A. and {Chartab}, Nima and {Hemmati}, Shoubaneh and {Long}, Arianna S. and {Sanjaripour}, Sogol},
        title = "{Searching for Quiescent Galaxies over $3 < z < 6$ in JWST Surveys Using Manifold Learning}",
      journal = {arXiv e-prints},
         year = 2025,
        month = jan,
          eid = {arXiv:2501.09066},
        pages = {arXiv:2501.09066},
          doi = {10.48550/arXiv.2501.09066},
archivePrefix = {arXiv},
       eprint = {2501.09066},
 primaryClass = {astro-ph.GA},
       adsurl = {https://ui.adsabs.harvard.edu/abs/2025arXiv250109066D}
}

@ARTICLE{Pizzati+24,
       author = {{Pizzati}, Elia and {Hennawi}, Joseph F. and {Schaye}, Joop and {Schaller}, Matthieu and {Eilers}, Anna-Christina and {Wang}, Feige and {Frenk}, Carlos S. and {Elbers}, Willem and {Helly}, John C. and {Mackenzie}, Ruari and {Matthee}, Jorryt and {Bordoloi}, Rongmon and {Kashino}, Daichi and {Naidu}, Rohan P. and {Yue}, Minghao},
        title = "{A unified model for the clustering of quasars and galaxies at z {\ensuremath{\approx}} 6}",
      journal = {\mnras},
         year = 2024,
        month = nov,
       volume = {534},
       number = {4},
        pages = {3155-3175},
          doi = {10.1093/mnras/stae2307},
archivePrefix = {arXiv},
       eprint = {2403.12140},
 primaryClass = {astro-ph.GA},
       adsurl = {https://ui.adsabs.harvard.edu/abs/2024MNRAS.534.3155P}
}

@ARTICLE{Xiao+24,
       author = {{Xiao}, Mengyuan and {Oesch}, Pascal A. and {Elbaz}, David and {Bing}, Longji and {Nelson}, Erica J. and {Weibel}, Andrea and {Illingworth}, Garth D. and {van Dokkum}, Pieter and {Naidu}, Rohan P. and {Daddi}, Emanuele and {Bouwens}, Rychard J. and {Matthee}, Jorryt and {Wuyts}, Stijn and {Chisholm}, John and {Brammer}, Gabriel and {Dickinson}, Mark and {Magnelli}, Benjamin and {Leroy}, Lucas and {Schaerer}, Daniel and {Herard-Demanche}, Thomas and {Lim}, Seunghwan and {Barrufet}, Laia and {Endsley}, Ryan and {Fudamoto}, Yoshinobu and {G{\'o}mez-Guijarro}, Carlos and {Gottumukkala}, Rashmi and {Labb{\'e}}, Ivo and {Magee}, Dan and {Marchesini}, Danilo and {Maseda}, Michael and {Qin}, Yuxiang and {Reddy}, Naveen A. and {Shapley}, Alice and {Shivaei}, Irene and {Shuntov}, Marko and {Stefanon}, Mauro and {Whitaker}, Katherine E. and {Wyithe}, J. Stuart B.},
        title = "{Accelerated formation of ultra-massive galaxies in the first billion years}",
      journal = {\nat},
         year = 2024,
        month = nov,
       volume = {635},
       number = {8038},
        pages = {311-315},
          doi = {10.1038/s41586-024-08094-5},
archivePrefix = {arXiv},
       eprint = {2309.02492},
 primaryClass = {astro-ph.GA},
       adsurl = {https://ui.adsabs.harvard.edu/abs/2024Natur.635..311X}
}

@ARTICLE{Carnall+24,
       author = {{Carnall}, A.~C. and {Cullen}, F. and {McLure}, R.~J. and {McLeod}, D.~J. and {Begley}, R. and {Donnan}, C.~T. and {Dunlop}, J.~S. and {Shapley}, A.~E. and {Rowlands}, K. and {Almaini}, O. and {Arellano-C{\'o}rdova}, K.~Z. and {Barrufet}, L. and {Cimatti}, A. and {Ellis}, R.~S. and {Grogin}, N.~A. and {Hamadouche}, M.~L. and {Illingworth}, G.~D. and {Koekemoer}, A.~M. and {Leung}, H. -H. and {Lovell}, C.~C. and {P{\'e}rez-Gonz{\'a}lez}, P.~G. and {Santini}, P. and {Stanton}, T.~M. and {Wild}, V.},
        title = "{The JWST EXCELS survey: too much, too young, too fast? Ultra-massive quiescent galaxies at 3 < z < 5}",
      journal = {\mnras},
         year = 2024,
        month = oct,
       volume = {534},
       number = {1},
        pages = {325-348},
          doi = {10.1093/mnras/stae2092},
archivePrefix = {arXiv},
       eprint = {2405.02242},
 primaryClass = {astro-ph.GA},
       adsurl = {https://ui.adsabs.harvard.edu/abs/2024MNRAS.534..325C}
}

@ARTICLE{Long+24b,
       author = {{Long}, Arianna S. and {Casey}, Caitlin M. and {McKinney}, Jed and {Zavala}, Jorge A. and {Akins}, Hollis B. and {Cooper}, Olivia R. and {Lambrides}, Matthieu Bethermin Erini L. and {Franco}, Maximilien and {Caputi}, Karina and {Champagne}, Jaclyn B. and {Man}, Allison W.~S. and {Treister}, Ezequiel and {Manning}, Sinclaire M. and {Sanders}, David B. and {Talia}, Margherita and {Aravena}, Manuel and {Clements}, D.~L. and {da Cunha}, Elisabete and {Faisst}, Andreas L. and {Gentile}, Fabrizio and {Hodge}, Jacqueline and {Brammer}, Gabriel and {Brusa}, Marcella and {Finkelstein}, Steven L. and {Fujimoto}, Seiji and {Hayward}, Christopher C. and {Ilbert}, Olivier and {Jolly}, Jean-Baptiste and {Kartaltepe}, Jeyhan S. and {Knudsen}, Kirsten and {Koekemoer}, Anton M. and {Liu}, Daizhong and {Magdis}, Georgios and {McCracken}, Henry Joy and {Rhodes}, Jason and {Robertson}, Brant E. and {Scoville}, Nick and {Sheth}, Kartik and {Smolcic}, Vernesa and {Spilker}, Justin and {Taniguchi}, Yoshiaki and {Toft}, Sune and {Urry}, C. Megan and {Yun}, Min},
        title = "{The Extended Mapping Obscuration to Reionization with ALMA (Ex-MORA) Survey: 5$\sigma$ Source Catalog and Redshift Distribution}",
      journal = {arXiv e-prints},
         year = 2024,
        month = aug,
          eid = {arXiv:2408.14546},
        pages = {arXiv:2408.14546},
          doi = {10.48550/arXiv.2408.14546},
archivePrefix = {arXiv},
       eprint = {2408.14546},
 primaryClass = {astro-ph.GA},
       adsurl = {https://ui.adsabs.harvard.edu/abs/2024arXiv240814546L}
}

@ARTICLE{Gao+24,
       author = {{Gao}, Zhen-Kai and {Lim}, Chen-Fatt and {Wang}, Wei-Hao and {Chen}, Chian-Chou and {Smail}, Ian and {Chapman}, Scott C. and {Zheng}, Xian Zhong and {Shim}, Hyunjin and {Kodama}, Tadayuki and {Ao}, Yiping and {Chang}, Siou-Yu and {Clements}, David L. and {Dunlop}, James S. and {Ho}, Luis C. and {Hsu}, Yun-Hsin and {Hwang}, Chorng-Yuan and {Hwang}, Ho Seong and {Koprowski}, M.~P. and {Scott}, Douglas and {Serjeant}, Stephen and {Toba}, Yoshiki and {Urquhart}, Sheona A.},
        title = "{SCUBA-2 Ultra Deep Imaging EAO Survey (STUDIES). V. Confusion-limited Submillimeter Galaxy Number Counts at 450 {\ensuremath{\mu}}m and Data Release for the COSMOS Field}",
      journal = {\apj},
         year = 2024,
        month = aug,
       volume = {971},
       number = {1},
          eid = {117},
        pages = {117},
          doi = {10.3847/1538-4357/ad53c1},
archivePrefix = {arXiv},
       eprint = {2405.20616},
 primaryClass = {astro-ph.GA},
       adsurl = {https://ui.adsabs.harvard.edu/abs/2024ApJ...971..117G}
}

@ARTICLE{Tanaka+24,
       author = {{Tanaka}, Masayuki and {Onodera}, Masato and {Shimakawa}, Rhythm and {Ito}, Kei and {Kakimoto}, Takumi and {Kubo}, Mariko and {Morishita}, Takahiro and {Toft}, Sune and {Valentino}, Francesco and {Wu}, Po-Feng},
        title = "{A Protocluster of Massive Quiescent Galaxies at z = 4}",
      journal = {\apj},
         year = 2024,
        month = jul,
       volume = {970},
       number = {1},
          eid = {59},
        pages = {59},
          doi = {10.3847/1538-4357/ad5316},
archivePrefix = {arXiv},
       eprint = {2311.11569},
 primaryClass = {astro-ph.GA},
       adsurl = {https://ui.adsabs.harvard.edu/abs/2024ApJ...970...59T}
}

@ARTICLE{Long+24a,
       author = {{Long}, Arianna S. and {Antwi-Danso}, Jacqueline and {Lambrides}, Erini L. and {Lovell}, Christopher C. and {de la Vega}, Alexander and {Valentino}, Francesco and {Zavala}, Jorge A. and {Casey}, Caitlin M. and {Wilkins}, Stephen M. and {Yung}, L.~Y. Aaron and {Arrabal Haro}, Pablo and {Bagley}, Micaela B. and {Bisigello}, Laura and {Chworowsky}, Katherine and {Cooper}, M.~C. and {Cooper}, Olivia R. and {Cooray}, Asantha R. and {Croton}, Darren and {Dickinson}, Mark and {Finkelstein}, Steven L. and {Franco}, Maximilien and {Gould}, Katriona M.~L. and {Hirschmann}, Michaela and {Hutchison}, Taylor A. and {Kartaltepe}, Jeyhan S. and {Kocevski}, Dale D. and {Koekemoer}, Anton M. and {Lucas}, Ray A. and {McKinney}, Jed and {Nere}, Rachel and {Papovich}, Casey and {P{\'e}rez-Gonz{\'a}lez}, Pablo G. and {Pirzkal}, Nor and {Santini}, Paola},
        title = "{Efficient NIRCam Selection of Quiescent Galaxies at 3 < z < 6 in CEERS}",
      journal = {\apj},
         year = 2024,
        month = jul,
       volume = {970},
       number = {1},
          eid = {68},
        pages = {68},
          doi = {10.3847/1538-4357/ad4cea},
archivePrefix = {arXiv},
       eprint = {2305.04662},
 primaryClass = {astro-ph.GA},
       adsurl = {https://ui.adsabs.harvard.edu/abs/2024ApJ...970...68L}
}

@ARTICLE{Perez-Gonzalez+24,
       author = {{P{\'e}rez-Gonz{\'a}lez}, Pablo G. and {Rinaldi}, Pierluigi and {Caputi}, Karina I. and {{\'A}lvarez-M{\'a}rquez}, Javier and {Annunziatella}, Marianna and {Langeroodi}, Danial and {Moutard}, Thibaud and {Boogaard}, Leindert and {Iani}, Edoardo and {Melinder}, Jens and {Costantin}, Luca and {{\"O}stlin}, G{\"o}ran and {Colina}, Luis and {Greve}, Thomas R. and {Wright}, Gillian and {Alonso-Herrero}, Almudena and {Bik}, Arjan and {Bosman}, Sarah E.~I. and {Crespo G{\'o}mez}, Alejandro and {Dicken}, Daniel and {Eckart}, Andreas and {Garc{\'\i}a-Mar{\'\i}n}, Macarena and {Gillman}, Steven and {G{\"u}del}, Manuel and {Henning}, Thomas and {Hjorth}, Jens and {Jermann}, Iris and {Labiano}, {\'A}lvaro and {Meyer}, Romain A. and {Pei{\ensuremath{\beta}}ker}, Florian and {Pye}, John P. and {Ray}, Thomas P. and {Tikkanen}, Tuomo and {Walter}, Fabian and {van der Werf}, Paul P.},
        title = "{A NIRCam-dark Galaxy Detected with the MIRI/F1000W Filter in the MIDIS/JADES Hubble Ultra Deep Field}",
      journal = {\apjl},
         year = 2024,
        month = jul,
       volume = {969},
       number = {1},
          eid = {L10},
        pages = {L10},
          doi = {10.3847/2041-8213/ad517b},
archivePrefix = {arXiv},
       eprint = {2402.16942},
 primaryClass = {astro-ph.GA},
       adsurl = {https://ui.adsabs.harvard.edu/abs/2024ApJ...969L..10P}
}








\noindent\rule{\columnwidth}{0.4pt}
$^{1}$Astronomy Centre, University of Sussex, Falmer, Brighton BN1 9QH, UK\\
$^{2}$Department of Astronomy, University of Geneva, Chemin Pegasi 51, 1290 Versoix, Switzerland\\
$^{3}$Aix-Marseille Universite, CNRS, CNES, LAM (Laboratoire d’Astrophysique de Marseille), Marseille, France\\
$^{4}$Argelander-Institut für Astronomie, Universität Bonn, Auf dem Hügel 71, 53121 Bonn, Germany\\
$^{5}$Purple Mountain Observatory, Chinese Academy of Sciences, 10 Yuanhua Road, Nanjing 210023, China\\
$^{6}$Universite Paris-Saclay, Universite Paris Cite, CEA, CNRS, AIM, 91191 Gif-sur-Yvette, France\\
$^{7}$Institut de Radioastronomie Millimétrique (IRAM), 300 rue la Piscine, 38406 Saint Martin d’Hères, France\\
$^{8}$Space Telescope Science Institute, 3700 San Martin Dr., Baltimore, MD 21218, USA\\
$^{9}$Cosmic Dawn Center (DAWN), Copenhagen, Denmark\\
$^{10}$DTU Space, Technical University of Denmark, Elektrovej 327, DK2800 Kgs. Lyngby, Denmark\\
$^{11}$Department for Astrophysical \& Planetary Science, University of Colorado, Boulder, CO 80309, USA\\
$^{12}$NSF Astronomy and Astrophysics Postdoctoral Fellow\\
$^{13}$IPAC, California Institute of Technology, 1200 E. California Blvd. Pasadena, CA 91125, USA\\
$^{14}$Department of Physics, University of California, Santa Barbara, Santa Barbara, CA 93106, USA\\
$^{15}$Laboratory for Multiwavelength Astrophysics, School of Physics and Astronomy, Rochester Institute of Technology, 84 Lomb Memorial Drive, Rochester, NY 14623, USA\\
$^{16}$Department of Astronomy, The University of Texas at Austin, 2515 Speedway Blvd Stop C1400, Austin, TX 78712, USA\\
$^{17}$Department of Astronomy, The University of Washington, Seattle, WA USA\\
$^{18}$Institut d’Astrophysique de Paris, UMR 7095, CNRS, and Sorbonne Université, 98 bis boulevard Arago, F-75014 Paris, France\\
$^{19}$Univ. Grenoble Alpes, CNRS, IPAG, 38000 Grenoble, France\\
$^{20}$Jet Propulsion Laboratory, California Institute of Technology, 4800 Oak Grove Drive, Pasadena, CA 91001, USA\\
$^{21}$Department of Astronomy and Astrophysics, University of California, Santa Cruz, 1156 High Street, Santa Cruz, CA 95064, USA\\
$^{22}$Institute for Astronomy, University of Hawaii, 2680 Woodlawn Drive, Honolulu, HI 96822, USA\\
$^{23}$Niels Bohr Institute, University of Copenhagen, Jagtvej 128, DK-2200, Copenhagen N, Denmark

\bsp	
\label{lastpage}
\end{document}